\documentclass[12pt]{article}
\usepackage{amsmath,amssymb,amsthm,amsxtra,
overpic,bbm,subfigure,float,color,bm,epsfig}
\def\thefootnote{\fnsymbol{footnote}}
\makeatletter
\renewcommand{\@thesubfigure}{\hskip\subfiglabelskip}
\makeatother

\begin{document}
\vspace{0.2cm}
	
\begin{center}
{\Large\bf 
Leptonic unitarity triangles: RGE running effects and
$\mu$-$\tau$ reflection symmetry breaking}
\end{center}	
\vspace{0.2cm}
	
\begin{center}
{\bf Jing-yu Zhu}\footnote{Email: zhujingyu@ihep.ac.cn} \\
{Institute of High Energy Physics, and School of Physical
Sciences, \\ University of Chinese Academy of Sciences, Beijing 
100049, China \\}
\end{center}

\vspace{1.5cm}
	
\begin{abstract}
There are six leptonic unitarity triangles (LUTs) defined by six 
orthogonality conditions
of the three-family lepton flavor mixing matrix in
the complex plane. In the 
framework of the standard model or the minimal supersymmetric 
standard model, the evolutions
of sides and inner angles of the six LUTs from a superhigh energy scale
$\Lambda_{\rm H}^{}$ to the electroweak scale 
$\Lambda_{\rm EW}^{}$
due to the renormalization-group equation (RGE) running
are derived in the integral form for both 
Dirac and Majorana neutrinos. Furthermore,
the LUTs as an intuitively geometrical language are applied to 
the description of
the RGE-induced $\mu$-$\tau$ reflection symmetry breaking
analytically and numerically.
\end{abstract}
	
\begin{flushleft}
\hspace{0.8cm} PACS number(s): 14.60.Pq, 11.30.Hv, 11.10.Hi.
\end{flushleft}

\def\thefootnote{\arabic{footnote}}
\setcounter{footnote}{0}	
\newpage
	
\section{Introduction}

In the recent twenty years, a series of neutrino oscillation
experiments have definitely proved that neutrinos have masses
and lepton flavors mix with one another \cite{PDG2018}. 
The latter can be described by the well-known 
Pontecorvo-Maki-Nakagawa-Sakata (PMNS) matrix $U$
\cite{PMNS1,PMNS2},
which connects three neutrino mass eigenstates 
$(\nu_1^{},\nu_2^{},\nu_3^{})$ and flavor
eigenstates $(\nu_e^{},\nu_{\mu}^{},\nu_{\tau}^{})$ by
\begin{eqnarray}
\begin{pmatrix}
\nu_e^{} \cr \nu_{\mu}^{} \cr \nu_{\tau}  \end{pmatrix}
\hspace{-0.17cm} & = &\hspace{-0.17cm}
\begin{pmatrix}
U_{e1}^{} & U_{e2}^{} & U_{e3}^{} \cr
U_{\mu 1}^{} & U_{\mu 2}^{} & U_{\mu 3}^{} \cr
U_{\tau 1}^{} & U_{\tau 2}^{} & U_{\tau 3}^{}
\end{pmatrix}
\begin{pmatrix}
\nu_1^{} \cr \nu_{2}^{} \cr \nu_{3}  \end{pmatrix} \;,
\end{eqnarray}
in the basis where the mass eigenstates of three charged leptons
are identical with their flavor eigenstates. 
According to the orthogonality of the rows and
columns of $U$, one may define six
leptonic unitarity triangles (LUTs) in the complex
plane as a geometrical 
language to intuitively describe lepton flavor mixing and CP 
violation \cite{Fritzsch1999}. 
The six triangles are
\begin{eqnarray}
&&\triangle_{e}^{} :  U_{\mu 1}^{} U_{\tau 1}^{*} + 
U_{\mu 2}^{} U_{\tau 2}^{*} + U_{\mu 3}^{} U_{\tau3}^{*}
 =  0 \;,\nonumber\\
&&\triangle_{\mu}^{} :  U_{\tau 1}^{} U_{e 1}^{*} + 
U_{\tau 2}^{} U_{e 2}^{*} + U_{\tau 3}^{} U_{e 3}^{*}
=  0 \;,\nonumber\\
&&\triangle_{\tau}^{} :  U_{e 1}^{} U_{\mu 1}^{*} + 
U_{e 2}^{} U_{\mu 2}^{*} + U_{e 3}^{} U_{\mu3}^{*}
=  0 \;,
\end{eqnarray}
which are insensitive to the Majorana phases; and 
\begin{eqnarray}
&&\triangle_{1}^{} :  U_{e 2}^{} U_{e 3}^{*} + 
U_{\mu 2}^{} U_{\mu 3}^{*} + U_{\tau 2}^{} U_{\tau 3}^{*}
=  0 \;,\nonumber\\
&&\triangle_{2}^{} :  U_{e 3}^{} U_{e 1}^{*} + 
U_{\mu 3}^{} U_{\mu 1}^{*} + U_{\tau 3}^{} U_{\tau 1}^{*}
=  0 \;,\nonumber\\
&&\triangle_{3}^{} :  U_{e 1}^{} U_{e 2}^{*} + 
U_{\mu 1}^{} U_{\mu 2}^{*} + U_{\tau 1}^{} U_{\tau 2}^{*}
=  0 \;,
\end{eqnarray}
whose orientations are fixed by the Majorana phases
\footnote{In the definitions of six LUTs in Eqs. (2) and (3), 
we do not consider the unphysical phases of $U$ 
for both Dirac and Majorana neutrinos. The orientations of
LUTs correspond to possible rotations caused by unphysical
or Majorana phases of $U$. So only the orientations of 
$\triangle_i^{}$ of
Majorana neutrinos have physical meaning.}
\cite{AguilarSaavedra:2000vr}.
The areas of these LUTs are all equal to $|{\cal J}|/2$, 
where ${\cal J}$ means the Jarlskog invariant of $U$ 
describing leptonic CP violation in neutrino oscillations 
and can be defined through
\begin{eqnarray}
{\rm Im}\left(U_{\alpha i}^{} U_{\beta i}^{*}
U_{\alpha j}^{*} U_{\beta j}^{}\right)=
{\cal J}\sum_{\gamma}^{} \epsilon_{\alpha\beta\gamma}^{}
\sum_{k}^{} \epsilon_{ijk}^{}.
\end{eqnarray}
The subscripts $(\alpha,\beta,\gamma)$ and $(i,j,k)$ in this 
paper always run over $(e,\mu,\tau)$ and $(1,2,3)$, respectively,
if not otherwise specified.
The six LUTs consist of eighteen vector sides in the
complex plane shown 
in Eqs. (2) and (3)
and nine inner angles which can be expressed as 
\begin{eqnarray}
\phi_{\alpha i}^{}  \hspace{-0.17cm} & \equiv &\hspace{-0.17cm}
\eta_{\phi}^{}{\rm arg} \left[- \frac{U_{\beta j}^{} U_{\gamma j}^{*}}
{U_{\beta k}^{} U_{\gamma k}^{*}}\right] =
\eta_{\phi}^{}{\rm arg} \left[- \frac{U_{\beta j}^{} U_{\beta k}^{*}}
{U_{\gamma j}^{} U_{\gamma k}^{*}}\right]\;.
\end{eqnarray} 
In Eq. (5), $\alpha$, $\beta$ and $\gamma$ run cyclically over
$e$, $\mu$ and $\tau$; $i$, $j$ and $k$ run
cyclically over $1$, $2$ and $3$;
 $\eta_{\phi}^{}=1$ for ${\cal J}<0$ and $\eta_{\phi}^{}=-1$ for ${\cal J}>0$
\footnote{We add $\eta_{\phi}^{}$ 
to ensure that the inner angles are
positive. In the relevant references, $\eta_{\phi}^{}$
was often neglected. }. 

The language of LUTs has been discussed in a number of
papers \cite{UT1,UT2,UT3,He,XingZhu2015,XingZhang,Luo} 
since it was
introduced into the lepton sector.
These papers mainly focus on the following aspects:
\begin{itemize}
\item
The reconstruction of LUTs through future precision
neutrino oscillation and non-oscillation experiments
will be a useful and intuitive geometric way to demonstrate CP 
violation in the lepton sector, and 
this will be complementary to the direct
measurements of CP asymmetries \cite{UT1,UT2}. 
Furthermore, testing whether the LUTs
are close will provide tests of the unitarity of 
the PMNS matrix, which might be violated 
due to the existence of sterile neutrinos
\cite{Fritzsch1999,AguilarSaavedra:2000vr,UT3}. 
\item
One can directly use the sides and inner angles of the LUTs to
describe neutrino-neutrino oscillations, 
neutrino-antineutrino oscillations and neutrino decays, 
where the inner angles of the LUTs 
have definite physical meaning \cite{He,XingZhu2015}. 
The shapes of 
the LUTs can be reformed  either by terrestrial matter effects,
or by renormalization-group-equation (RGE) running effects, 
or by some other new physics effects, implying the corrections 
of such effects to lepton flavor mixing and CP violation
\cite{XingZhang,Luo}. There are
also discussions about the underlying phenomenological meaning of
special shapes of the LUTs \cite{XingZhu2015}.
\end{itemize}
In Ref. \cite{Luo}, the RGE running behaviors of 
inner angles of the LUTs have been discussed 
in the differential form.
In this paper, we aim to study how the sides and inner angles 
of the LUTs evolve in the integral form 
due to the RGE running from an arbitrary 
superhigh energy scale $\Lambda_{\rm H}^{}$ to the electroweak
scale $\Lambda_{\rm EW}^{}$ in
the framework of the standard model (SM) or the
minimal supersymmetric standard model (MSSM). Both the cases 
of Dirac and Majorana neutrinos will be consideblack. 
We get the RGE-induced corrections to the LUTs
by performing perturbative expansions. The final
analytical results are
independent of the parametrization of $U$. Assuming
the $\mu$-$\tau$ reflection symmetry \cite{Harrison:2002et,Xing:2015fdg}
to be satisfied at a superhigh energy $\Lambda_{\mu\tau}^{}$,
the corresponding $\triangle_i^{}$ should be isosceles triangles;
$\triangle_{\mu}^{}$ and $\triangle_{\tau}^{}$
are congruent with each other.
When running down to $\Lambda_{\rm EW}^{}$, the $\mu$-$\tau$ reflection 
symmetry will be broken due the RGE running effects, leading to
the deviations of the LUTs from their special shapes at $\Lambda_{\mu\tau}^{}$.
So the corrections to the LUTs from $\Lambda_{\mu\tau}^{}$
to $\Lambda_{\rm EW}^{}$ can be used to intuitively describe 
the corresponding
RGE-induced $\mu$-$\tau$ reflection symmetry breaking, and
thus it is meaningful to explore how the LUTs can be reformed 
analytically and numerically in this case.

The rest of this paper is organized as follows. 
In section 2, we derive the RGE-induced connections of the sides 
and inner angles of the LUTs between $\Lambda_{\rm EW}^{}$
and $\Lambda_{\rm H}^{}$ in the integral form
in the framework of the SM or the MSSM, where both Dirac and 
Majorana neutrinos are consideblack.
Section 3 is devoted to simplifying the analytically approximate 
results in section 2 by assuming the $\mu$-$\tau$ reflection symmetry
at $\Lambda_{\mu\tau}^{}$. In section 4, the RGE-induced deviations of the LUTs
from the $\mu$-$\tau$ reflection symmetry limits
will be studied numerically by scanning the complete parameter
space, where the smallest neutrino mass and the MSSM parameter $\tan\beta$ 
at $\Lambda_{\rm EW}^{}$ run in the 
reasonable ranges $\left[0,0.1\right]$ eV and $\left[10, 
50\right]$, respectively, just as the way taken in Ref. \cite{Huang}. The normal
mass ordering (NMO) and inverted mass ordering 
(IMO) of Dirac
or Majorana neutrinos will be consideblack. Finally, section
5 is a summary of our main results.
\section{RGE running effects on the LUTs}
\subsection{The case of Dirac neutrinos}
Before a decisive measurement of the neutrinoless 
double-beta decay \cite{Furry:1939qr}
verifies the Majorana nature of massive neutrinos, 
it is meaningful to consider the cases of both Dirac 
and Majorana neutrinos theoretically \cite{D1}.
The evolution of the Dirac neutrino mass matrix
from $\Lambda_{\rm H}^{}$ to
$\Lambda_{\rm EW}^{}$ in the integral form can be written as
\cite{Xing:2017mkx}
\begin{eqnarray}
M_{\nu}^{\prime} \hspace{-0.17cm} & 
= &\hspace{-0.17cm} I_0^{} T_{l}^{}  M_{\nu}^{} \;,
\end{eqnarray}
where $M_{\nu}^{}$ and $M_{\nu}^{\prime}$ are the Dirac neutrino mass
matrix at $\Lambda_{\rm H}^{}$ and $\Lambda_{\rm EW}^{}$,
respectively. Note that the notations with a prime superscript 
in the following text
denote the parameters at $\Lambda_{\rm EW}^{}$ and those without
such a superscript
stand for the corresponding parameters 
at $\Lambda_{\rm H}^{}$ if not otherwise specified.
Here we define $T_l^{}= {{\rm Diag} \{I_e, I_{\mu}^{}, I_{\tau}^{}\}}$ 
and
\begin{eqnarray}
I_0^{} \hspace{-0.17cm} & = &\hspace{-0.17cm}
{\rm exp} \left[\frac{1}{16 \pi^2}
\int^{0}_{t^{\prime}}  G \ {\rm d} t\right] \;,\nonumber\\
I_{\alpha}^{} \hspace{-0.17cm} & = &\hspace{-0.17cm}
{\rm exp} \left[- \frac{C_l^{}}{16 \pi^2} \int^{0}_{t^{\prime}}
y_{\alpha}^2 \ {\rm d} t\right]\;,
\end{eqnarray}
where $t \equiv{\rm ln} \left(\mu/\Lambda_{\rm H}^{}\right)$ 
with $\mu$ being an arbitrary renormalization scale between 
$\Lambda_{\rm EW}^{}$
and $\Lambda_{\rm H}^{}$, and $t^{\prime} \equiv{\rm ln} \left(\Lambda_{\rm 
EW}^{}/\Lambda_{\rm H}^{}\right)$. In the SM, one has
$G=0.45 g_1^2 + 2.25 g_2^2 -  3 y_t^2$
and $C_l^{}=-1.5$;
while in the MSSM, $G=0.6 g_1^2 + 3 g_2^2 -  3 y_t^2$
and $C_l^{}=1$, where $g_{1,2}^{}$
dnote the gauge couplings, and $y_t^{}$ and $y_{\alpha}^{}$ (for 
$\alpha = e,\mu,\tau$) stand respectively
for the Yukawa couplings of the top quark and charged leptons
\footnote{Eqs. (6) and (21)
can also apply to the two-Higgs-doublet
models (2HDMs) \cite{Branco:2011iw} 
only by replacing the definitions
of $G$ and $C_l^{}$. For example, in type-II 2HDM,
$C_l^{}=-1.5$ and $G=0.45 g_1^2 +2.25 g_2^2 - y_t^2- 3 y_b^2$
with $y_b^{}$ being the bottom-quark Yukawa coupling
\cite{Antusch:2005gp}.}. 
 The Hermitian matrix $H_{\nu}^{}=
M_{\nu}^{} M_{\nu}^{\dagger}$ can be diagonalized by  the 
unitary transfomation $U^{\dagger}H_{\nu}^{}
U ={\rm Diag} \{{m_1^2, m_2^2,m_3^2 } \}$ with $m_i^{}$ being
the neutrino masses. Similarly we have $H_{\nu}^{\prime}=
M_{\nu}^{\prime} M_{\nu}^{\prime\dagger}$ and
$U^{\prime\dagger}H_{\nu}^{\prime} U^{\prime}={\rm Diag} \{{m_1^{\prime 
2}, m_2^{\prime 2},m_3^{\prime 2} } \}$
at $\Lambda_{\rm EW}^{}$. According to Eq. (6), we
directly write out
\begin{eqnarray}
\left(H_{\nu}^{\prime}\right)_{\alpha\beta} 
\hspace{-0.17cm} & = &\hspace{-0.17cm}
\sum_{i}^{} m_i^{\prime 2} U_{\alpha i}^{\prime} 
U_{\beta i}^{\prime *} =I_0^2 I_{\alpha}^{} I_{\beta}^{}
\left(H_{\nu}^{}\right)_{\alpha\beta}^{} \nonumber\;,\\
\left(H_{\nu}^{\prime} H_{\nu}^{\prime\dagger}\right)_{\alpha\beta} 
\hspace{-0.17cm} & = &\hspace{-0.17cm}
\sum_{i}^{} m_i^{\prime 4} U_{\alpha i}^{\prime} 
U_{\beta i}^{\prime *} =I_0^4 I_{\alpha}^{} I_{\beta}^{}
\sum_{\gamma}^{} I_{\gamma}^2\left(H_{\nu}^{}\right)_{\alpha\gamma}^{}
\left(H_{\nu}^{}\right)_{\gamma\beta}^{} \;.
\end{eqnarray}
Then Eq. (8), together with
the unitarity conditions of $U^{\prime}$, 
can yield a full set of linear equations of $U_{\alpha i}^{\prime}
U_{\beta i}^{\prime *}$
\begin{eqnarray}
\begin{pmatrix}
1 & 1 & 1\cr m_1^{\prime 2} & m_2^{\prime 2} & m_3^{\prime 2} \cr
m_1^{\prime 4} & m_2^{\prime 4} & m_3^{\prime 4} 
\end{pmatrix}
\begin{pmatrix}U_{\alpha 1}^{\prime}U_{\beta 1}^{\prime *} \cr
U_{\alpha 2}^{\prime}U_{\beta 2}^{\prime *} \cr
U_{\alpha 3}^{\prime}U_{\beta 3}^{\prime *} \cr
\end{pmatrix}=\begin{pmatrix}\delta_{\alpha\beta} \cr
I_0^2 I_{\alpha}^{} I_{\beta} \left(H_{\nu}^{}\right)_{\alpha\beta} \cr
I_0^4 I_{\alpha}^{} I_{\beta}^{}\sum \limits_{\gamma} 
I_{\gamma}^2\left(H_{\nu}^{}\right)_{\alpha\gamma}
\left(H_{\nu}^{}\right)_{\gamma\beta}
\end{pmatrix}
\end{eqnarray}
from which we can get exact expressions of $U_{\alpha i}^{\prime}
U_{\beta i}^{\prime *}$ at $\Lambda_{\rm EW}^{}$. 
In addition, calculating the determinnant of $H_{\nu}^{\prime}$
and the traces of $H_{\nu}^{\prime}$ and $H_{\nu}^{\prime} 
H_{\nu}^{\prime\dagger}$ from Eq. (8) leads to
\begin{eqnarray}
m_1^{\prime 2} m_2^{\prime 2} m_3^{\prime 2}
\hspace{-0.17cm} & = &\hspace{-0.17cm}
I_0^6 I_e^2 I_{\mu}^2 I_{\tau}^2 m_1^2 m_2^2 m_3^2 \nonumber \;,\\
m_1^{\prime 2} + m_2^{\prime 2} + m_3^{\prime 2}
\hspace{-0.17cm} & = &\hspace{-0.17cm}
I_0^2\sum_{\alpha}^{} I_{\alpha}^2 \sum_{i}^{} m_i^2 \left|U^{}
_{\alpha i}\right|^2 \nonumber \;,\\
m_1^{\prime 4} + m_2^{\prime 4} + m_3^{\prime 4}
\hspace{-0.17cm} & = &\hspace{-0.17cm} I_0^4
\sum_{\alpha}^{} I_{\alpha}^2 \sum^{}_{\beta} I_{\beta}^2
\left|\sum_i^{} m_i^2 U_{\alpha i}^{} U_{\beta i}^{*} \right|^2 \;.
\end{eqnarray}
One can see $m_i^{\prime 2}$ (for $i =1,2,3$
and $m_1^{\prime 2}\ne m_2^{\prime 2} \ne m_3^{\prime 2} $) should be the
solutions of the equation of $\lambda$:
\begin{eqnarray}
\lambda^3 - b \lambda^2 + \frac{b^2 -c}{2} \lambda -a =0 \;,
\end{eqnarray}
with $a = m_1^{\prime 2}  m_2^{\prime 2}  m_3^{\prime 2}$,
$b= m_1^{\prime 2}+ m_2^{\prime 2}+ m_3^{\prime 2}$
and  $c=m_1^{\prime 4} + m_2^{\prime 4} + m_3^{\prime 4}$
coming from Eq. (10). 
The exact but complicated expressions of 
$m_i^{\prime 2}$ have been shown in appendix A. 
Here we perform some analytical approximations of Eq. (11)
to see more clearly the dependence of $m_i^{\prime 2}$ 
on the parameters at $\Lambda_{\rm H}^{}$.
The tau-dominance approximation $T_l^{}\simeq
{\rm Diag} \{1,1, \left(1 -\epsilon\right)\}$ 
will be taken due to the relationship
$y_e^2 \ll y_{\mu}^{2} \ll y_{\tau}^2$ with
\begin{eqnarray}
\epsilon \hspace{-0.17cm} & = &\hspace{-0.17cm}
\frac{C_l^{}}{16 \pi^2} \int^0_{t^{\prime}} y_{\tau}^2 {\rm d} t \;,
\end{eqnarray}
which is a small quantity, at most of order ${\cal O}(0.01)$
\cite{Huang}.
Expanding Eq. (11) in $\epsilon$ up to the first order gives rise to
\begin{eqnarray}
m_1^{\prime 2} \hspace{-0.17cm} & \simeq &\hspace{-0.17cm}
I_0^2 m_1^2 \left(1 - 2 \epsilon |U_{\tau 1}^{}|^2 \right) 
\;,\nonumber\\
m_2^{\prime 2} \hspace{-0.17cm} & \simeq &\hspace{-0.17cm}
I_0^2 m_2^2 \left(1 - 2 \epsilon |U_{\tau 2}^{}|^2 \right) 
\;,\nonumber\\
m_3^{\prime 2} \hspace{-0.17cm} & \simeq &\hspace{-0.17cm}
I_0^2 m_3^2 \left(1 - 2 \epsilon |U_{\tau 3}^{}|^2 \right) 
\;.
\end{eqnarray}
By inserting the tau-dominance approximation of 
$T_l^{}$ and the above approximate results of $m_i^{\prime 2}$ 
into Eq. (9), and
expanding it in $\epsilon$ up to the first order,
we can arrive at the analytical approximations of 
$|U_{\alpha i}^{\prime}|^2$ at $\Lambda_{\rm EW}^{}$:   
\begin{flalign}
&|U_{e1}^{\prime}|^2   \simeq 
|U_{e1}^{}|^2 - \frac{2 \epsilon}{\Delta_{21}^{} \Delta_{31}^{}}
\left[\left(m_2^2 m_3^2 -m_1^4\right) |U_{e1}^{}|^2 |U_{\tau 1}^{}|^2
+ m_1^2  \Delta_{32}^{} \left(|U_{e2}^{}|^2 |U_{\tau 2}^{}|^2
-|U_{e 3}^{}|^2 |U_{\tau 3}^{}|^2\right)\right]\;, &\nonumber \\
&|U_{e2}^{\prime}|^2 \simeq 
|U_{e2}^{}|^2 + \frac{2 \epsilon}{\Delta_{21}^{} \Delta_{32}^{}}
\left[\left(m_1^2 m_3^2 -m_2^4\right) |U_{e2}^{}|^2 |U_{\tau 2}^{}|^2
+ m_2^2  \Delta_{31}^{} \left(|U_{e1}^{}|^2 |U_{\tau 1}^{}|^2
-|U_{e 3}^{}|^2 |U_{\tau 3}^{}|^2\right)\right] \;, \nonumber \\
&|U_{e3}^{\prime}|^2  \simeq 
|U_{e3}^{}|^2 - \frac{2 \epsilon}{\Delta_{31}^{} \Delta_{32}^{}}
\left[\left(m_1^2 m_2^2 -m_3^4\right) |U_{e3}^{}|^2 |U_{\tau 3}^{}|^2
+ m_3^2  \Delta_{21}^{} \left(|U_{e1}^{}|^2 |U_{\tau 1}^{}|^2
-|U_{e 2}^{}|^2 |U_{\tau 2}^{}|^2\right)\right] \;; &
\end{flalign}
and
\begin{flalign}
&|U_{\mu 1}^{\prime}|^2  \simeq 
|U_{\mu 1}^{}|^2 - \frac{2 \epsilon}{\Delta_{21}^{} \Delta_{31}^{}}
\left[\left(m_2^2 m_3^2 -m_1^4\right) |U_{\mu 1}^{}|^2 
|U_{\tau 1}^{}|^2
+ m_1^2  \Delta_{32}^{} \left(|U_{\mu 2}^{}|^2 |U_{\tau 2}^{}|^2
-|U_{\mu 3}^{}|^2 |U_{\tau 3}^{}|^2\right)\right]\;,& \nonumber \\
&|U_{\mu 2}^{\prime}|^2  \simeq 
|U_{\mu 2}^{}|^2 + \frac{2 \epsilon}{\Delta_{21}^{} \Delta_{32}^{}}
\left[\left(m_1^2 m_3^2 -m_2^4\right) |U_{\mu 2}^{}|^2 
|U_{\tau 2}^{}|^2
+ m_2^2  \Delta_{31}^{} \left(|U_{\mu 1}^{}|^2 |U_{\tau 1}^{}|^2
-|U_{\mu 3}^{}|^2 |U_{\tau 3}^{}|^2\right)\right] \;,& \nonumber \\
&|U_{\mu 3}^{\prime}|^2   \simeq
|U_{\mu 3}^{}|^2 - \frac{2 \epsilon}{\Delta_{31}^{} \Delta_{32}^{}}
\left[\left(m_1^2 m_2^2 -m_3^4\right) |U_{\mu 3}^{}|^2 
|U_{\tau 3}^{}|^2
+ m_3^2  \Delta_{21}^{} \left(|U_{\mu 1}^{}|^2 |U_{\tau 1}^{}|^2
-|U_{\mu 2}^{}|^2 |U_{\tau 2}^{}|^2\right)\right] \;; 
\end{flalign}		
and
\begin{eqnarray}
|U_{\tau 1}^{\prime}|^2 \hspace{-0.17cm} & \simeq &\hspace{-0.17cm}
|U_{\tau 1}^{}|^2 + \frac{2 \epsilon |U_{\tau 1}^{}|^2}
{\Delta_{21}^{} \Delta_{31}^{}}
\left[\left(m_2^2 m_3^2 -m_1^4\right) \left(1 -
|U_{\tau 1}^{}|^2 \right)
+ m_1^2  \Delta_{32}^{} \left( |U_{\tau 2}^{}|^2
- |U_{\tau 3}^{}|^2\right)\right]\;, \nonumber \\
|U_{\tau 2}^{\prime}|^2 \hspace{-0.17cm} & \simeq &\hspace{-0.17cm}
|U_{\tau 2}^{}|^2 - \frac{2 \epsilon |U_{\tau 2}^{}|^2}
{\Delta_{21}^{} \Delta_{32}^{}}
\left[\left(m_1^2 m_3^2 -m_2^4\right) \left(1 -
|U_{\tau 2}^{}|^2 \right)
+ m_2^2  \Delta_{31}^{} \left( |U_{\tau 1}^{}|^2
- |U_{\tau 3}^{}|^2\right)\right]\;, \nonumber \\
|U_{\tau 3}^{\prime}|^2 \hspace{-0.17cm} & \simeq &\hspace{-0.17cm}
|U_{\tau 3}^{}|^2 + \frac{2 \epsilon |U_{\tau 3}^{}|^2}
{\Delta_{31}^{} \Delta_{32}^{}}
\left[\left(m_1^2 m_2^2 -m_3^4\right) \left(1 -
|U_{\tau 3}^{}|^2 \right)
+ m_3^2  \Delta_{21}^{} \left( |U_{\tau 1}^{}|^2
- |U_{\tau 2}^{}|^2\right)\right]\;,
\end{eqnarray}
where $\Delta_{21}^{} = m_2^{2} -m_1^2$, $\Delta_{31}^{} = 
m_3^{2} -m_1^2$ and $\Delta_{32}^{} = m_3^{2} -m_2^2$. 
One can see that $|U_{\alpha i}^{\prime}|^2$ depend on
$|U_{\tau i}^{}|^2$ besides $|U_{\alpha i}^{}|^2$ 
owing to the tau-dominance approximation of $T_l^{}$,
and there will be similar characteristics in the following results.
The analytical approximations of $|U_{\alpha i}^{\prime}|^2$ 
in Eqs. (14)---(16)
satisfy $\sum_i |U_{\alpha i}^{\prime}|^2= \sum_{\alpha} 
|U_{\alpha i}^{\prime}|^2=1$.
In the same way, we can obtain the analytical approximations 
of the vector sides of 
$\triangle_{\alpha}^{\prime}$ at $\Lambda_{\rm EW}^{}$:
\begin{eqnarray}
U_{\mu 1}^{\prime} U_{\tau 1}^{\prime *} \hspace{-0.17cm} & \simeq 
&\hspace{-0.17cm} U_{\mu 1}^{} U_{\tau 1}^{*} + \frac{
	\epsilon}{\Delta_{21}^{} \Delta_{31}^{}} \left\{\left[\left(
m_2^2 m_3^2 - m_1^4 \right) \left(1- 2 |U_{\tau 1}|^2
\right) - m_1^{2} \Delta_{32}^{} \left(1- 2 |U_{\tau 2}|^2
\right)\right] U_{\mu 1}^{} U_{\tau 1}^{*} 
\right. \nonumber\\&&  \left.\hspace{-0.17cm}
- 2 m_1^2 \Delta_{32}^{} 
|U_{\tau 1}|^2 U_{\mu 3}^{} U_{\tau 3}^{*}
\right\} \; , \nonumber\\
U_{\mu 2}^{\prime} U_{\tau 2}^{\prime *} \hspace{-0.17cm} & \simeq 
&\hspace{-0.17cm} U_{\mu 2}^{} U_{\tau 2}^{*} - \frac{
	\epsilon}{\Delta_{21}^{} \Delta_{32}^{}} \left\{\left[\left(
m_1^2 m_3^2 - m_2^4 \right) \left(1- 2 |U_{\tau 2}|^2
\right) - m_2^{2} \Delta_{31}^{} \left(1- 2 |U_{\tau 1}|^2
\right)\right] U_{\mu 2}^{} U_{\tau 2}^{*} 
\right. \nonumber\\&&  \left.\hspace{-0.17cm}
- 2 m_2^2 \Delta_{31}^{} 
|U_{\tau 2}|^2 U_{\mu 3}^{} U_{\tau 3}^{*} 
\right\} \; , \nonumber\\
U_{\mu 3}^{\prime} U_{\tau 3}^{\prime *} \hspace{-0.17cm} & \simeq 
&\hspace{-0.17cm} U_{\mu 3}^{} U_{\tau 3}^{*} + \frac{
	\epsilon}{\Delta_{31}^{} \Delta_{32}^{}} \left\{\left[\left(
m_1^2 m_2^2 - m_3^4 \right) \left(1- 2 |U_{\tau 3}|^2
\right) + m_3^{2} \Delta_{21}^{} \left(1- 2 |U_{\tau 2}|^2
\right)\right] U_{\mu 3}^{} U_{\tau 3}^{*} 
\right. \nonumber\\&&  \left.\hspace{-0.17cm}
+ 2 m_3^2 \Delta_{21}^{} 
|U_{\tau 3}|^2 U_{\mu 1}^{} U_{\tau 1}^{*}
\right\} \;,  
\end{eqnarray}
for $\triangle_e^{\prime}$; and
\begin{eqnarray}
U_{\tau 1}^{\prime} U_{e 1}^{\prime *} \hspace{-0.17cm} & \simeq 
&\hspace{-0.17cm} U_{\tau 1}^{} U_{e 1}^{*} + \frac{
	\epsilon}{\Delta_{21}^{} \Delta_{31}^{}} \left\{\left[\left(
m_2^2 m_3^2 - m_1^4 \right) \left(1- 2 |U_{\tau 1}|^2
\right) - m_1^{2} \Delta_{32}^{} \left(1- 2 |U_{\tau 2}|^2
\right)\right] U_{\tau 1}^{} U_{e 1}^{*} 
\right. \nonumber\\&&  \left.\hspace{-0.17cm}
- 2 m_1^2 \Delta_{32}^{} 
|U_{\tau 1}|^2 U_{\tau 3}^{} U_{e 3}^{*}
\right\} \; , \nonumber\\
U_{\tau 2}^{\prime} U_{e 2}^{\prime *} \hspace{-0.17cm} & \simeq 
&\hspace{-0.17cm} U_{\tau 2}^{} U_{e 2}^{*} - \frac{
	\epsilon}{\Delta_{21}^{} \Delta_{32}^{}} \left\{\left[\left(
m_1^2 m_3^2 - m_2^4 \right) \left(1- 2 |U_{\tau 2}|^2
\right) - m_2^{2} \Delta_{31}^{} \left(1- 2 |U_{\tau 1}|^2
\right)\right] U_{\tau 2}^{} U_{e 2}^{*} 
\right. \nonumber\\&&  \left.\hspace{-0.17cm}
- 2 m_2^2 \Delta_{31}^{} 
|U_{\tau 2}|^2 U_{\tau 3}^{} U_{e 3}^{*}
\right\} \; , \nonumber\\
U_{\tau 3}^{\prime} U_{e 3}^{\prime *} \hspace{-0.17cm} & \simeq 
&\hspace{-0.17cm} U_{\tau 3}^{} U_{e 3}^{*} + \frac{
	\epsilon}{\Delta_{31}^{} \Delta_{32}^{}} \left\{\left[\left(
m_1^2 m_2^2 - m_3^4 \right) \left(1- 2 |U_{\tau 3}|^2
\right) + m_3^{2} \Delta_{21}^{} \left(1- 2 |U_{\tau 2}|^2
\right)\right] U_{\tau 3}^{} U_{e 3}^{*} 
\right. \nonumber\\&&  \left.\hspace{-0.17cm}
+ 2 m_3^2 \Delta_{21}^{} 
|U_{\tau 3}|^2 U_{\tau 1}^{} U_{e 1}^{*} 
\right\} \;, 
\end{eqnarray}
for $\triangle_{\mu}^{\prime}$; and
\begin{eqnarray}
U_{e 1}^{\prime} U_{\mu 1}^{\prime *} \hspace{-0.17cm} & \simeq 
&\hspace{-0.17cm} U_{e 1}^{} U_{\mu 1}^{*} - \frac{
	2 \epsilon}{\Delta_{21}^{} \Delta_{31}^{}} \left\{\left[\left(
m_2^2 m_3^2 - m_1^4 \right) |U_{\tau 1}|^2
- m_1^{2} \Delta_{32}^{} |U_{\tau 2}|^2
\right] U_{e 1}^{} U_{\mu 1}^{*} 
\right. \nonumber\\&&  \left.\hspace{-0.17cm}
-  m_1^2 \Delta_{32}^{} 
\left(1 -|U_{\tau 1}|^2 \right) U_{e 3}^{} U_{\mu 3}^{*}
\right\} \; , \nonumber\\
U_{e 2}^{\prime} U_{\mu 2}^{\prime *} \hspace{-0.17cm} & \simeq 
&\hspace{-0.17cm} U_{e 2}^{} U_{\mu 2}^{*} + \frac{
	2 \epsilon}{\Delta_{21}^{} \Delta_{32}^{}} \left\{\left[\left(
m_1^2 m_3^2 - m_2^4 \right) |U_{\tau 2}|^2
- m_2^{2} \Delta_{31}^{} |U_{\tau 1}|^2
\right] U_{e 2}^{} U_{\mu 2}^{*} 
\right. \nonumber\\&&  \left.\hspace{-0.17cm}
-  m_2^2 \Delta_{31}^{} 
\left(1 -|U_{\tau 2}|^2 \right) U_{e 3}^{} U_{\mu 3}^{*} 
\right\} \; , \nonumber\\
U_{e 3}^{\prime} U_{\mu 3}^{\prime *} \hspace{-0.17cm} & \simeq 
&\hspace{-0.17cm} U_{e 3}^{} U_{\mu 3}^{*} - \frac{
	2 \epsilon}{\Delta_{31}^{} \Delta_{32}^{}} \left\{\left[\left(
m_1^2 m_2^2 - m_3^4 \right) |U_{\tau 3}|^2
+ m_3^{2} \Delta_{21}^{} |U_{\tau 2}|^2
\right] U_{e 3}^{} U_{\mu 3}^{*} 
\right. \nonumber\\&&  \left.\hspace{-0.17cm}
+  m_3^2 \Delta_{21}^{} 
\left(1 -|U_{\tau 3}|^2 \right) U_{e 1}^{} U_{\mu 1}^{*} 
\right\} \;, 
\end{eqnarray}
for $\triangle_{\tau}^{\prime}$, where
$\sum_i^{} U_{\alpha i}^{\prime}U_{\beta i}^{\prime*}=0$
holds for $(\alpha,\beta)= (e, \mu), (\mu,\tau), (\tau, e)$.
Considering the fact that the lengths of three sides 
of each $\Delta_{i}^{\prime}$ can be derived from 
Eqs. (14)---(16), one can see that the six LUTs 
$\left(\triangle_{\alpha}^{\prime}\right.$ 
and $\left.\triangle_{i}^{\prime}\right)$  of the
Dirac neutrinos at $\Lambda_{\rm EW}^{}$ can be 
approximately fixed from the above results. Furthermore,
we can get the approximate Jarlskog 
invariant ${\cal J}^{\prime}$ of the Dirac neutrinos 
at $\Lambda_{\rm EW}^{}$ from anyone of Eqs. (17)---(19). 
The result is
\begin{eqnarray}
{\cal J}^{\prime} \hspace{-0.17cm} & \simeq &\hspace{-0.17cm}
{\cal J} - \frac{2 \epsilon {\cal J}}{\Delta_{21}^{}
	\Delta_{31}^{} \Delta_{32}^{} } \left[ m_1^{2}
\left( m_2^{4} + m_3^{4}\right) \left(|U_{\tau 3}^{}|^2
-|U_{\tau 2}^{}|^2 \right) + m_2^{2}
\left( m_1^{4} + m_3^{4}\right) \left(|U_{\tau 1}^{}|^2
-|U_{\tau 3}^{}|^2 \right) \right.\nonumber\\&&\hspace{-0.17cm}
\left. + m_3^{2}
\left( m_1^{4} + m_2^{4}\right) \left(|U_{\tau 2}^{}|^2
-|U_{\tau 1}^{}|^2 \right)
\right] \;.
\end{eqnarray}
With the help of Eqs. (4) and (5), one can define
\begin{eqnarray}
\cot \phi_{\alpha i}^{} \hspace{-0.17cm} & \equiv &\hspace{-0.17cm}
\frac{\eta_{\phi}^{}}{\cal J} {\rm Re} \left(U_{\beta j}^{} 
U_{\gamma j}^{*} U_{\beta k}^{*} U_{\gamma k}^{}\right) \;, \nonumber\\
\cot \phi_{\alpha i}^{\prime} \hspace{-0.17cm} & \equiv &\hspace{-0.17cm}
\frac{\eta_{\phi}^{}}{\cal J^{\prime}} {\rm Re} \left(U_{\beta j}^{\prime} 
U_{\gamma j}^{\prime*} U_{\beta k}^{\prime*} U_{\gamma 
	k}^{\prime}\right) \;,
\end{eqnarray}
where $(\alpha,\beta,\gamma)$ and $(i,j,k)$ run cyclically
over $(e,\mu,\tau)$ and $(1,2,3)$, respectively.
We then calculate the evolutions of 
the nine inner angles of LUTs for the Dirac neutrinos from 
$\Lambda_{\rm H}^{}$ to $\Lambda_{\rm EW}^{}$ by 
combining Eqs. (17)---(19) and Eq. (21), and obtain
\begin{eqnarray}
\cot \phi_{e1}^{\prime} \hspace{-0.17cm} & \simeq &\hspace{-0.17cm}
\cot \phi_{e1}^{} + \eta_{\phi}^{} \frac{2 \epsilon |U_{\tau 2}^{}|^2 |U_{\tau 3}^{}|^2
}{{\cal J} \Delta_{21}^{} \Delta_{31}^{} \Delta_{32}^{}}
\left(m_2^{2} \Delta_{31}^{2} |U_{\mu 3}^{}|^2 -
m_3^{2} \Delta_{21}^{2} |U_{\mu 2}^{}|^2\right) \nonumber \;,\\
\cot \phi_{e2}^{\prime} \hspace{-0.17cm} & \simeq &\hspace{-0.17cm}
\cot \phi_{e2}^{} + \eta_{\phi}^{}
\frac{2 \epsilon |U_{\tau 1}^{}|^2 |U_{\tau 3}^{}|^2
}{{\cal J} \Delta_{21}^{} \Delta_{31}^{} \Delta_{32}^{}}
\left(m_3^{2} \Delta_{21}^{2} |U_{\mu 1}^{}|^2 -
m_1^{2} \Delta_{32}^{2} |U_{\mu 3}^{}|^2 \right) \nonumber \;,\\
\cot \phi_{e3}^{\prime} \hspace{-0.17cm} & \simeq &\hspace{-0.17cm}
\cot \phi_{e3}^{} + \eta_{\phi}^{}\frac{2 \epsilon |U_{\tau 1}^{}|^2 |U_{\tau 2}^{}|^2
}{{\cal J} \Delta_{21}^{} \Delta_{31}^{} \Delta_{32}^{}}
\left(m_1^{2} \Delta_{32}^{2} |U_{\mu 2}^{}|^2 -
m_2^{2} \Delta_{31}^{2} |U_{\mu 1}^{}|^2 \right)  \; ;
\end{eqnarray}
and
\begin{eqnarray}
\cot \phi_{\mu 1}^{\prime} \hspace{-0.17cm} & \simeq &\hspace{-0.17cm}
\cot \phi_{\mu 1}^{} + \eta_{\phi}^{}
\frac{2 \epsilon |U_{\tau 2}^{}|^2 |U_{\tau 3}^{}|^2
}{{\cal J} \Delta_{21}^{} \Delta_{31}^{} \Delta_{32}^{}}
\left(m_2^{2} \Delta_{31}^{2} |U_{e 3}^{}|^2 -
m_3^{2} \Delta_{21}^{2} |U_{e 2}^{}|^2\right) \nonumber \;,\\
\cot \phi_{\mu 2}^{\prime} \hspace{-0.17cm} & \simeq &\hspace{-0.17cm}
\cot \phi_{\mu 2}^{} + \eta_{\phi}^{}
\frac{2 \epsilon |U_{\tau 1}^{}|^2 |U_{\tau 3}^{}|^2
}{{\cal J} \Delta_{21}^{} \Delta_{31}^{} \Delta_{32}^{}}
\left(m_3^{2} \Delta_{21}^{2} |U_{e 1}^{}|^2 -
m_1^{2} \Delta_{32}^{2} |U_{e 3}^{}|^2 \right) \nonumber \;,\\
\cot \phi_{\mu 3}^{\prime} \hspace{-0.17cm} & \simeq &\hspace{-0.17cm}
\cot \phi_{\mu 3}^{} + \eta_{\phi}^{}
\frac{2 \epsilon |U_{\tau 1}^{}|^2 |U_{\tau 2}^{}|^2
}{{\cal J} \Delta_{21}^{} \Delta_{31}^{} \Delta_{32}^{}}
\left(m_1^{2} \Delta_{32}^{2} |U_{e 2}^{}|^2 -
m_2^{2} \Delta_{31}^{2} |U_{e 1}^{}|^2 \right)  \; ;
\end{eqnarray}
and
\begin{flalign}
&\cot \phi_{\tau 1}^{\prime}   \simeq 
\cot \phi_{\tau 1}^{} + \frac{2 \eta_{\phi}^{}\epsilon
}{{\cal J} \Delta_{21}^{} \Delta_{31}^{} \Delta_{32}^{}}
\left[m_3^{2} \Delta_{21}^{2} |U_{e 2}^{} U_{\mu 2}^{*}|^2 
\left( 1- |U_{\tau 3}^{}|^2\right) -
m_2^{2} \Delta_{31}^{2} |U_{e 3}^{} U_{\mu 3}^{*}|^2
\left( 1- |U_{\tau 2}^{}|^2\right)\right] \nonumber \;,&\\
&\cot \phi_{\tau 2}^{\prime}  \simeq 
\cot \phi_{\tau 2}^{} + \frac{2 \eta_{\phi}^{}\epsilon
}{{\cal J} \Delta_{21}^{} \Delta_{31}^{} \Delta_{32}^{}}
\left[m_1^{2} \Delta_{32}^{2} |U_{e 3}^{} U_{\mu 3}^{*}|^2 
\left( 1- |U_{\tau 1}^{}|^2\right) -
m_3^{2} \Delta_{21}^{2} |U_{e 1}^{} U_{\mu 1}^{*}|^2
\left( 1- |U_{\tau 3}^{}|^2\right)\right] \nonumber \;,&\\
&\cot \phi_{\tau 3}^{\prime}  \simeq 
\cot \phi_{\tau 3}^{} + \frac{2 \eta_{\phi}^{}\epsilon
}{{\cal J} \Delta_{21}^{} \Delta_{31}^{} \Delta_{32}^{}}
\left[m_2^{2} \Delta_{31}^{2} |U_{e 1}^{} U_{\mu 1}^{*}|^2 
\left( 1- |U_{\tau 2}^{}|^2\right) -
m_1^{2} \Delta_{32}^{2} |U_{e 2}^{} U_{\mu 2}^{*}|^2
\left( 1- |U_{\tau 1}^{}|^2\right)\right]  \; .&
\end{flalign}
\subsection{The case of Majorana neutrinos}
When considering the Majorana neutrinos, 
one can naturally explain their small masses 
through the seesaw mechanisms \cite {SS}. 
The evolution of the Majorana neutrino mass matrix
from $\Lambda_{\rm H}^{}$ to $\Lambda_{\rm EW}^{}$ 
in the integral form  can be written as 
\cite{Fritzsch1999,Ellis:1999my} 
\begin{eqnarray}
M_{\nu}^{\prime} \hspace{-0.17cm} & 
= &\hspace{-0.17cm} I_0^{2} T_{l}^{}  M_{\nu}^{} T_{l}^{}\;.
\end{eqnarray}
Note that $M_{\nu}^{}$ and $M_{\nu}^{\prime}$ represent
the Majorana neutrino mass matrices at $\Lambda_{\rm H}^{}$ 
and $\Lambda_{\rm EW}^{}$, respectively.
$I_0^{}$ and $T_l^{}$
have been defined below Eq. (6) and
in Eq. (7). We can also derive the direct 
connections of the LUTs between the two energy scales 
$\Lambda_{\rm H}^{}$ and $\Lambda_{\rm EW}^{}$ as in 
the Dirac case.
Let us repeat the similar calculations below for comparison. 
We first diagonalize $M_{\nu}^{}$ and $M_{\nu}^{\prime}$
through $U^{\dagger}_{} M_{\nu}^{}U^{*}_{}={\rm Diag}
\{m_1^2,m_2^2,m_3^2\}$ and $U^{\prime\dagger}_{} 
M_{\nu}^{\prime}U^{\prime*}_{}={\rm Diag}
\{m_1^{\prime 2},m_2^{\prime 2},m_3^{\prime 2}\}$.
According to Eq. (25), the Hermitian matrices $H_{\nu}^{\prime}
\equiv M_{\nu}^{} M_{\nu}^{\dagger}$
and $H_{\nu}^{\prime}H_{\nu}^{\prime\dagger}$ for 
the case of Majorana neutrinos can be expressed as
\begin{eqnarray}
\left(H_{\nu}^{\prime}\right)_{\alpha\beta} 
\hspace{-0.17cm} & = &\hspace{-0.17cm}
\sum_{i}^{} m_i^{\prime 2} U_{\alpha i}^{\prime} 
U_{\beta i}^{\prime *} =I_0^4 I_{\alpha}^{} I_{\beta}^{}
Q_{\alpha\beta}^{}\nonumber\;,\\
\left(H_{\nu}^{\prime} H_{\nu}^{\prime\dagger}\right)_{\alpha\beta} 
\hspace{-0.17cm} & = &\hspace{-0.17cm}
\sum_{i}^{} m_i^{\prime 4} U_{\alpha i}^{\prime} 
U_{\beta i}^{\prime *} =I_0^8 I_{\alpha}^{} I_{\beta}^{}
\sum_{\gamma}^{} I_{\gamma}^2 Q_{\alpha\gamma}^{}
Q_{\gamma\beta}^{} \;,
\end{eqnarray}
where we have defined $Q \equiv M_{\nu}^{} T_l^2 
M_{\nu}^{\dagger}$
for simplicity. With the help of Eq. (26) and
the unitarity conditions of $U^{\prime}$, 
we can get
\begin{eqnarray}
\begin{pmatrix}
1 & 1 & 1\cr m_1^{\prime 2} & m_2^{\prime 2} & m_3^{\prime 2} \cr
m_1^{\prime 4} & m_2^{\prime 4} & m_3^{\prime 4} 
\end{pmatrix}
\begin{pmatrix}U_{\alpha 1}^{\prime}U_{\beta 1}^{\prime *} \cr
U_{\alpha 2}^{\prime}U_{\beta 2}^{\prime *} \cr
U_{\alpha 3}^{\prime}U_{\beta 3}^{\prime *} \cr
\end{pmatrix}=\begin{pmatrix}\delta_{\alpha\beta} \cr
I_0^4 I_{\alpha}^{} I_{\beta} Q_{\alpha\beta}^{} \cr
I_0^8 I_{\alpha}^{} I_{\beta}^{}\sum \limits_{\gamma} 
I_{\gamma}^2 Q_{\alpha\gamma}^{}
Q_{\gamma\beta}^{}
\end{pmatrix}.
\end{eqnarray}
Moreover, the determinant of $H_{\nu}^{\prime}$
together with the traces of $H_{\nu}^{\prime}$ and $H_{\nu}^{\prime} 
H_{\nu}^{\prime\dagger}$ leads to
\begin{eqnarray}
m_1^{\prime 2} m_2^{\prime 2} m_3^{\prime 2}
\hspace{-0.17cm} & = &\hspace{-0.17cm}
I_0^{12} I_e^4 I_{\mu}^4 I_{\tau}^4 m_1^2 m_2^2 m_3^2 \nonumber \;,\\
m_1^{\prime 2} + m_2^{\prime 2} + m_3^{\prime 2}
\hspace{-0.17cm} & = &\hspace{-0.17cm}
I_0^4\sum_{\alpha}^{} I_{\alpha}^2 
\sum_{\beta}^{} I_{\beta}^2\left|\left(M_{\nu}^{}\right)
_{\alpha\beta}^{} \right|^2\nonumber \;,\\
m_1^{\prime 4} + m_2^{\prime 4} + m_3^{\prime 4}
\hspace{-0.17cm} & = &\hspace{-0.17cm} I_0^8
\sum_{\alpha}^{} I_{\alpha}^2 \sum^{}_{\beta} I_{\beta}^2
\left|\sum_{\gamma}^{} I_{\gamma}^2 \left(M_{\nu}^{}\right)
_{\alpha \gamma}^{} \left(M_{\nu}^{}\right)
_{\beta \gamma}^{*}\right|^2 \;,
\end{eqnarray}
where $\left(M_{\nu}^{}\right)_{\alpha\beta}^{}=\sum_i^{}
m_i^{} U_{\alpha i}^{}U_{\beta i}^{}$.
By solving the equation
\begin{eqnarray}
\lambda^3 - b \lambda^2 + \frac{b^2 -c}{2} \lambda -a =0 \;,
\end{eqnarray}
with $a = m_1^{\prime 2}  m_2^{\prime 2}  m_3^{\prime 2}$,
$b= m_1^{\prime 2}+ m_2^{\prime 2}+ m_3^{\prime 2}$
and  $c=m_1^{\prime 4} + m_2^{\prime 4} + m_3^{\prime 4}$
coming from Eq. (28),
the exact expressions of $m_i^{\prime 2}$ for Majorana neutrinos
can be derived. 
One may refer to appendix A for their specific expressions. 
Here we calculate $m_i^{\prime 2}$ approximately by
expanding Eq. (29) in $\epsilon$, and arrive at
\begin{eqnarray}
m_1^{\prime 2} \hspace{-0.17cm} & \simeq &\hspace{-0.17cm}
I_0^4 m_1^2 \left(1 - 4 \epsilon |U_{\tau 1}^{}|^2 \right) 
\;,\nonumber\\
m_2^{\prime 2} \hspace{-0.17cm} & \simeq &\hspace{-0.17cm}
I_0^4 m_2^2 \left(1 - 4 \epsilon |U_{\tau 2}^{}|^2 \right) 
\;,\nonumber\\
m_3^{\prime 2} \hspace{-0.17cm} & \simeq &\hspace{-0.17cm}
I_0^4 m_3^2 \left(1 - 4 \epsilon |U_{\tau 3}^{}|^2 \right) 
\;.
\end{eqnarray}
By inserting the tau-dominance approximation of 
$T_l^{}$ and Eq. (30) into Eq. (27), 
and expanding it in $\epsilon$ up to the 
first order,
we can get the analytical approximations of 
$|U_{\alpha i}^{\prime}|^2$ at $\Lambda_{\rm EW}^{}$:   
\begin{eqnarray}
|U_{e1}^{\prime}|^2 \hspace{-0.17cm} & \simeq &\hspace{-0.17cm}
|U_{e1}^{}|^2 - \frac{2 \epsilon}{\Delta_{21}^{} \Delta_{31}^{}}
\left[\left(m_2^2 m_3^2 -m_1^4\right) |U_{e1}^{}|^2 |U_{\tau 1}^{}|^2
+ m_1^2  \Delta_{32}^{} \left(|U_{e2}^{}|^2 |U_{\tau 2}^{}|^2
-|U_{e 3}^{}|^2 |U_{\tau 3}^{}|^2\right)\right.\nonumber\\&&\left.
\hspace{-0.17cm} - 2 m_1^{} m_2^{} \Delta_{31}^{}{\mathbb R}_{e	
\tau}^{12}- 2 m_1^{} m_3^{} \Delta_{21}^{}
{\mathbb R}_{e \tau}^{13}\right]\;, \nonumber \\
|U_{e2}^{\prime}|^2 \hspace{-0.17cm} & \simeq &\hspace{-0.17cm}
|U_{e2}^{}|^2 + \frac{2 \epsilon}{\Delta_{21}^{} \Delta_{32}^{}}
\left[\left(m_1^2 m_3^2 -m_2^4\right) |U_{e2}^{}|^2 |U_{\tau 2}^{}|^2
+ m_2^2  \Delta_{31}^{} \left(|U_{e1}^{}|^2 |U_{\tau 1}^{}|^2
-|U_{e 3}^{}|^2 |U_{\tau 3}^{}|^2\right)\right.\nonumber\\&&\left.
\hspace{-0.17cm} - 2 m_1^{} m_2^{} \Delta_{32}^{}
{\mathbb R}_{e \tau}^{12} + 2 m_2^{} m_3^{} \Delta_{21}^{}
{\mathbb R}_{e \tau}^{23}\right]\;, \nonumber \\
|U_{e3}^{\prime}|^2 \hspace{-0.17cm} & \simeq &\hspace{-0.17cm}
|U_{e3}^{}|^2 - \frac{2 \epsilon}{\Delta_{31}^{} \Delta_{32}^{}}
\left[\left(m_1^2 m_2^2 -m_3^4\right) |U_{e3}^{}|^2 |U_{\tau 3}^{}|^2
- m_3^2  \Delta_{21}^{} \left(|U_{e2}^{}|^2 |U_{\tau 2}^{}|^2
-|U_{e 1}^{}|^2 |U_{\tau 1}^{}|^2\right)\right.\nonumber\\&&\left.
\hspace{-0.17cm} + 2 m_1^{} m_3^{} \Delta_{32}^{}
{\mathbb R}_{e \tau}^{13} + 2 m_2^{} m_3^{} \Delta_{31}^{}
{\mathbb R}_{e \tau}^{23}\right]\;;
\end{eqnarray}
and
\begin{eqnarray}
|U_{\mu 1}^{\prime}|^2 \hspace{-0.17cm} & \simeq &\hspace{-0.17cm}
|U_{\mu 1}^{}|^2 - \frac{2 \epsilon}{\Delta_{21}^{} \Delta_{31}^{}}
\left[\left(m_2^2 m_3^2 -m_1^4\right) |U_{\mu 1}^{}|^2 |U_{\tau 1}^{}|^2
+ m_1^2  \Delta_{32}^{} \left(|U_{\mu 2}^{}|^2 |U_{\tau 2}^{}|^2
-|U_{\mu 3}^{}|^2 |U_{\tau 3}^{}|^2\right)\right.\nonumber\\&&\left.
\hspace{-0.17cm} - 2 m_1^{} m_2^{} \Delta_{31}^{}
{\mathbb R}_{\mu \tau}^{12}- 2 m_1^{} m_3^{} \Delta_{21}^{}
{\mathbb R}_{\mu \tau}^{13}\right]\;, \nonumber \\
|U_{\mu 2}^{\prime}|^2 \hspace{-0.17cm} & \simeq &\hspace{-0.17cm}
|U_{\mu 2}^{}|^2 + \frac{2 \epsilon}{\Delta_{21}^{} \Delta_{32}^{}}
\left[\left(m_1^2 m_3^2 -m_2^4\right) |U_{\mu 2}^{}|^2 |U_{\tau 2}^{}|^2
+ m_2^2  \Delta_{31}^{} \left(|U_{\mu 1}^{}|^2 |U_{\tau 1}^{}|^2
-|U_{\mu 3}^{}|^2 |U_{\tau 3}^{}|^2\right)\right.\nonumber\\&&\left.
\hspace{-0.17cm} - 2 m_1^{} m_2^{} \Delta_{32}^{}
{\mathbb R}_{\mu \tau}^{12} + 2 m_2^{} m_3^{} \Delta_{21}^{}
{\mathbb R}_{\mu \tau}^{23}\right]\;, \nonumber \\
|U_{\mu 3}^{\prime}|^2 \hspace{-0.17cm} & \simeq &\hspace{-0.17cm}
|U_{\mu 3}^{}|^2 - \frac{2 \epsilon}{\Delta_{31}^{} \Delta_{32}^{}}
\left[\left(m_1^2 m_2^2 -m_3^4\right) |U_{\mu 3}^{}|^2 |U_{\tau 3}^{}|^2
- m_3^2  \Delta_{21}^{} \left(|U_{\mu 2}^{}|^2 |U_{\tau 2}^{}|^2
-|U_{\mu 1}^{}|^2 |U_{\tau 1}^{}|^2\right)\right.\nonumber\\&&\left.
\hspace{-0.17cm} + 2 m_1^{} m_3^{} \Delta_{32}^{}
{\mathbb R}_{\mu \tau}^{13} + 2 m_2^{} m_3^{} \Delta_{31}^{}
{\mathbb R}_{\mu \tau}^{23}\right]\;;
\end{eqnarray}
and
\begin{eqnarray}
|U_{\tau 1}^{\prime}|^2 \hspace{-0.17cm} & \simeq &\hspace{-0.17cm}
|U_{\tau 1}^{}|^2 + \frac{2 \epsilon}
{\Delta_{21}^{} \Delta_{31}^{}}
\left[\left(m_2^2 m_3^2 -m_1^4\right)  |U_{\tau 1}^{}|^2\left(1 -
|U_{\tau 1}^{}|^2 \right)
+ m_1^2  \Delta_{32}^{}  |U_{\tau 1}^{}|^2\left( |U_{\tau 2}^{}|^2
- |U_{\tau 3}^{}|^2\right) \right.\nonumber \\
&&\hspace{-0.17cm}\left.+ 2 m_1^{} m_2^{} \Delta_{31}^{}
{\mathbb R}_{\tau \tau}^{12} + 2 m_1^{} m_3^{} \Delta_{21}^{}
{\mathbb R}_{\tau \tau}^{13} \right]\;, \nonumber \\
|U_{\tau 2}^{\prime}|^2 \hspace{-0.17cm} & \simeq &\hspace{-0.17cm}
|U_{\tau 2}^{}|^2 - \frac{2 \epsilon}
{\Delta_{21}^{} \Delta_{32}^{}}
\left[\left(m_1^2 m_3^2 -m_2^4\right)  |U_{\tau 2}^{}|^2\left(1 -
|U_{\tau 2}^{}|^2 \right)
+ m_2^2  \Delta_{31}^{}  |U_{\tau 2}^{}|^2\left( |U_{\tau 1}^{}|^2
- |U_{\tau 3}^{}|^2\right) \right.\nonumber \\
&&\hspace{-0.17cm}\left.+ 2 m_1^{} m_2^{} \Delta_{32}^{}
{\mathbb R}_{\tau \tau}^{12}
- 2 m_2^{} m_3^{} \Delta_{21}^{} {\mathbb R}_{\tau \tau}^{23}
\right]\;, \nonumber \\
|U_{\tau 3}^{\prime}|^2 \hspace{-0.17cm} & \simeq &\hspace{-0.17cm}
|U_{\tau 3}^{}|^2 + \frac{2 \epsilon}
{\Delta_{31}^{} \Delta_{32}^{}}
\left[\left(m_1^2 m_2^2 -m_3^4\right)  |U_{\tau 3}^{}|^2\left(1 -
|U_{\tau 3}^{}|^2 \right)
- m_3^2  \Delta_{21}^{}  |U_{\tau 3}^{}|^2\left( |U_{\tau 2}^{}|^2
- |U_{\tau 1}^{}|^2\right) \right.\nonumber \\
&&\hspace{-0.17cm}\left.- 2 m_1^{} m_3^{} \Delta_{32}^{}
{\mathbb R}_{\tau \tau}^{13}
- 2 m_2^{} m_3^{} \Delta_{31}^{} {\mathbb R}_{\tau \tau}^{23}
\right]\;,
\end{eqnarray}
with ${\mathbb R}_{\alpha\beta}^{ij}$ denoting the
real parts of $U_{\alpha i}^{} U_{\alpha j}^{*}
U_{\beta i}^{} U_{\beta j}^{*}$. The vector sides of 
$\triangle_{\alpha}^{\prime}$ at $\Lambda_{\rm EW}^{}$
turn out to be:
\begin{eqnarray}
U_{\mu 1}^{\prime} U_{\tau 1}^{\prime *} \hspace{-0.17cm} & \simeq 
&\hspace{-0.17cm} U_{\mu 1}^{} U_{\tau 1}^{*} + \frac{
	\epsilon}{\Delta_{21}^{} \Delta_{31}^{}} \left\{\left[\left(
m_2^2 m_3^2 - m_1^4 \right) \left(1- 2 |U_{\tau 1}|^2
\right) - m_1^{2} \Delta_{32}^{} \left(1- 2 |U_{\tau 2}|^2
\right)\right] U_{\mu 1}^{} U_{\tau 1}^{*} 
\right. \nonumber\\&&  \left.\hspace{-0.17cm}
- 2 m_1^2 \Delta_{32}^{} 
|U_{\tau 1}|^2 U_{\mu 3}^{} U_{\tau 3}^{*} +
2 m_1^{} m_2^{} \Delta_{31}^{} \left(U_{\mu 1}^{}
U_{\tau 1}^{} U_{\tau 2}^{*2} + U_{\mu 2}^{}
U_{\tau 2}^{} U_{\tau 1}^{*2} \right) 
\right. \nonumber\\&&  \left.\hspace{-0.17cm} +
2 m_1^{} m_3^{} \Delta_{21}^{} \left(U_{\mu 1}^{}
U_{\tau 1}^{} U_{\tau 3}^{*2} + U_{\mu 3}^{}
U_{\tau 3}^{} U_{\tau 1}^{*2} \right)
\right\} \; , \nonumber\\
U_{\mu 2}^{\prime} U_{\tau 2}^{\prime *} \hspace{-0.17cm} & \simeq 
&\hspace{-0.17cm} U_{\mu 2}^{} U_{\tau 2}^{*} - \frac{
	\epsilon}{\Delta_{21}^{} \Delta_{32}^{}} \left\{\left[\left(
m_1^2 m_3^2 - m_2^4 \right) \left(1- 2 |U_{\tau 2}|^2
\right) - m_2^{2} \Delta_{31}^{} \left(1- 2 |U_{\tau 1}|^2
\right)\right] U_{\mu 2}^{} U_{\tau 2}^{*} 
\right. \nonumber\\&&  \left.\hspace{-0.17cm}
- 2 m_2^2 \Delta_{31}^{} 
|U_{\tau 2}|^2 U_{\mu 3}^{} U_{\tau 3}^{*} +
2 m_1^{} m_2^{} \Delta_{32}^{} \left(U_{\mu 1}^{}
U_{\tau 1}^{} U_{\tau 2}^{*2} + U_{\mu 2}^{}
U_{\tau 2}^{} U_{\tau 1}^{*2} \right) 
\right. \nonumber\\&&  \left.\hspace{-0.17cm} -
2 m_2^{} m_3^{} \Delta_{21}^{} \left(U_{\mu 2}^{}
U_{\tau 2}^{} U_{\tau 3}^{*2} + U_{\mu 3}^{}
U_{\tau 3}^{} U_{\tau 2}^{*2} \right)
\right\} \; , \nonumber\\
U_{\mu 3}^{\prime} U_{\tau 3}^{\prime *} \hspace{-0.17cm} & \simeq 
&\hspace{-0.17cm} U_{\mu 3}^{} U_{\tau 3}^{*} + \frac{
	\epsilon}{\Delta_{31}^{} \Delta_{32}^{}} \left\{\left[\left(
m_1^2 m_2^2 - m_3^4 \right) \left(1- 2 |U_{\tau 3}|^2
\right) + m_3^{2} \Delta_{21}^{} \left(1- 2 |U_{\tau 2}|^2
\right)\right] U_{\mu 3}^{} U_{\tau 3}^{*} 
\right. \nonumber\\&&  \left.\hspace{-0.17cm}
+ 2 m_3^2 \Delta_{21}^{} 
|U_{\tau 3}|^2 U_{\mu 1}^{} U_{\tau 1}^{*} -
2 m_2^{} m_3^{} \Delta_{31}^{} \left(U_{\mu 2}^{}
U_{\tau 2}^{} U_{\tau 3}^{*2} + U_{\mu 3}^{}
U_{\tau 3}^{} U_{\tau 2}^{*2} \right) 
\right. \nonumber\\&&  \left.\hspace{-0.17cm} -
2 m_1^{} m_3^{} \Delta_{32}^{} \left(U_{\mu 1}^{}
U_{\tau 1}^{} U_{\tau 3}^{*2} + U_{\mu 3}^{}
U_{\tau 3}^{} U_{\tau 1}^{*2} \right)
\right\} \;, 
\end{eqnarray}
for $\triangle_{e}^{\prime}$; and
\begin{eqnarray}
U_{\tau 1}^{\prime} U_{e 1}^{\prime *} \hspace{-0.17cm} & \simeq 
&\hspace{-0.17cm} U_{\tau 1}^{} U_{e 1}^{*} + \frac{
	\epsilon}{\Delta_{21}^{} \Delta_{31}^{}} \left\{\left[\left(
m_2^2 m_3^2 - m_1^4 \right) \left(1- 2 |U_{\tau 1}|^2
\right) - m_1^{2} \Delta_{32}^{} \left(1- 2 |U_{\tau 2}|^2
\right)\right] U_{\tau 1}^{} U_{e 1}^{*} 
\right. \nonumber\\&&  \left.\hspace{-0.17cm}
- 2 m_1^2 \Delta_{32}^{} 
|U_{\tau 1}|^2 U_{\tau 3}^{} U_{e 3}^{*} +
2 m_1^{} m_2^{} \Delta_{31}^{} \left(U_{e 1}^{*}
U_{\tau 1}^{*} U_{\tau 2}^{2} + U_{e 2}^{*}
U_{\tau 2}^{*} U_{\tau 1}^{2} \right) 
\right. \nonumber\\&&  \left.\hspace{-0.17cm} +
2 m_1^{} m_3^{} \Delta_{21}^{} \left(U_{e 1}^{*}
U_{\tau 1}^{*} U_{\tau 3}^{2} + U_{e 3}^{*}
U_{\tau 3}^{*} U_{\tau 1}^{2} \right)
\right\} \; , \nonumber\\
U_{\tau 2}^{\prime} U_{e 2}^{\prime *} \hspace{-0.17cm} & \simeq 
&\hspace{-0.17cm} U_{\tau 2}^{} U_{e 2}^{*} - \frac{
	\epsilon}{\Delta_{21}^{} \Delta_{32}^{}} \left\{\left[\left(
m_1^2 m_3^2 - m_2^4 \right) \left(1- 2 |U_{\tau 2}|^2
\right) - m_2^{2} \Delta_{31}^{} \left(1- 2 |U_{\tau 1}|^2
\right)\right] U_{\tau 2}^{} U_{e 2}^{*} 
\right. \nonumber\\&&  \left.\hspace{-0.17cm}
- 2 m_2^2 \Delta_{31}^{} 
|U_{\tau 2}|^2 U_{\tau 3}^{} U_{e 3}^{*} +
2 m_1^{} m_2^{} \Delta_{32}^{} \left(U_{e 1}^{*}
U_{\tau 1}^{*} U_{\tau 2}^{2} + U_{e 2}^{*}
U_{\tau 2}^{*} U_{\tau 1}^{2} \right) 
\right. \nonumber\\&&  \left.\hspace{-0.17cm} -
2 m_2^{} m_3^{} \Delta_{21}^{} \left(U_{e 2}^{*}
U_{\tau 2}^{*} U_{\tau 3}^{2} + U_{e 3}^{*}
U_{\tau 3}^{*} U_{\tau 2}^{2} \right)
\right\} \; , \nonumber\\
U_{\tau 3}^{\prime} U_{e 3}^{\prime *} \hspace{-0.17cm} & \simeq 
&\hspace{-0.17cm} U_{\tau 3}^{} U_{e 3}^{*} + \frac{
	\epsilon}{\Delta_{31}^{} \Delta_{32}^{}} \left\{\left[\left(
m_1^2 m_2^2 - m_3^4 \right) \left(1- 2 |U_{\tau 3}|^2
\right) + m_3^{2} \Delta_{21}^{} \left(1- 2 |U_{\tau 2}|^2
\right)\right] U_{\tau 3}^{} U_{e 3}^{*} 
\right. \nonumber\\&&  \left.\hspace{-0.17cm}
+ 2 m_3^2 \Delta_{21}^{} 
|U_{\tau 3}|^2 U_{\tau 1}^{} U_{e 1}^{*} -
2 m_2^{} m_3^{} \Delta_{31}^{} \left(U_{e 2}^{*}
U_{\tau 2}^{*} U_{\tau 3}^{2} + U_{e 3}^{*}
U_{\tau 3}^{*} U_{\tau 2}^{2} \right) 
\right. \nonumber\\&&  \left.\hspace{-0.17cm} -
2 m_1^{} m_3^{} \Delta_{32}^{} \left(U_{e 1}^{*}
U_{\tau 1}^{*} U_{\tau 3}^{2} + U_{e 3}^{*}
U_{\tau 3}^{*} U_{\tau 1}^{2} \right)
\right\} \;,  \nonumber\\
\end{eqnarray}
for $\triangle_{\mu}^{\prime}$; and
\begin{eqnarray}
U_{e 1}^{\prime} U_{\mu 1}^{\prime *} \hspace{-0.17cm} & \simeq 
&\hspace{-0.17cm} U_{e 1}^{} U_{\mu 1}^{*} - \frac{
2 \epsilon}{\Delta_{21}^{} \Delta_{31}^{}} \left\{\left[\left(
m_2^2 m_3^2 - m_1^4 \right) |U_{\tau 1}|^2
- m_1^{2} \Delta_{32}^{} |U_{\tau 2}|^2
\right] U_{e 1}^{} U_{\mu 1}^{*} 
\right. \nonumber\\&&  \left.\hspace{-0.17cm}
-  m_1^2 \Delta_{32}^{} 
\left(1 -|U_{\tau 1}|^2 \right) U_{e 3}^{} U_{\mu 3}^{*} -
m_1^{} m_2^{} \Delta_{31}^{} \left(U_{e 1}^{}
U_{\tau 1}^{} U_{\mu 2}^{*} U_{\tau 2}^{*} + 
U_{e 2}^{} U_{\tau 2}^{}
U_{\mu 1}^{*} U_{\tau 1}^{*} \right) 
\right. \nonumber\\&&  \left.\hspace{-0.17cm} -
m_1^{} m_3^{} \Delta_{21}^{} \left(U_{e 1}^{}
U_{\tau 1}^{} U_{\mu 3}^{*} U_{\tau 3}^{*} + U_{e 3}^{}
U_{\tau 3}^{} U_{\mu 1}^{*} U_{\tau 1}^{*} \right)
\right\} \; , \nonumber\\
U_{e 2}^{\prime} U_{\mu 2}^{\prime *} \hspace{-0.17cm} & \simeq 
&\hspace{-0.17cm} U_{e 2}^{} U_{\mu 2}^{*} + \frac{
2 \epsilon}{\Delta_{21}^{} \Delta_{32}^{}} \left\{\left[\left(
m_1^2 m_3^2 - m_2^4 \right) |U_{\tau 2}|^2
- m_2^{2} \Delta_{31}^{} |U_{\tau 1}|^2
\right] U_{e 2}^{} U_{\mu 2}^{*} 
\right. \nonumber\\&&  \left.\hspace{-0.17cm}
-  m_2^2 \Delta_{31}^{} 
\left(1 -|U_{\tau 2}|^2 \right) U_{e 3}^{} U_{\mu 3}^{*} -
m_1^{} m_2^{} \Delta_{32}^{} \left(U_{e 1}^{}
U_{\tau 1}^{} U_{\mu 2}^{*} U_{\tau 2}^{*} + 
U_{e 2}^{} U_{\tau 2}^{}
U_{\mu 1}^{*} U_{\tau 1}^{*} \right) 
\right. \nonumber\\&&  \left.\hspace{-0.17cm} +
m_2^{} m_3^{} \Delta_{21}^{} \left(U_{e 2}^{}
U_{\tau 2}^{} U_{\mu 3}^{*} U_{\tau 3}^{*} + U_{e 3}^{}
U_{\tau 3}^{} U_{\mu 2}^{*} U_{\tau 2}^{*} \right)
\right\} \; , \nonumber\\
U_{e 3}^{\prime} U_{\mu 3}^{\prime *} \hspace{-0.17cm} & \simeq 
&\hspace{-0.17cm} U_{e 3}^{} U_{\mu 3}^{*} - \frac{
	2 \epsilon}{\Delta_{31}^{} \Delta_{32}^{}} \left\{\left[\left(
m_1^2 m_2^2 - m_3^4 \right) |U_{\tau 3}|^2
+ m_3^{2} \Delta_{21}^{} |U_{\tau 2}|^2
\right] U_{e 3}^{} U_{\mu 3}^{*} 
\right. \nonumber\\&&  \left.\hspace{-0.17cm}
+  m_3^2 \Delta_{21}^{} 
\left(1 -|U_{\tau 3}|^2 \right) U_{e 1}^{} U_{\mu 1}^{*} +
m_1^{} m_3^{} \Delta_{32}^{} \left(U_{e 1}^{}
U_{\tau 1}^{} U_{\mu 3}^{*} U_{\tau 3}^{*} + 
U_{e 3}^{} U_{\tau 3}^{}
U_{\mu 1}^{*} U_{\tau 1}^{*} \right) 
\right. \nonumber\\&&  \left.\hspace{-0.17cm} +
m_2^{} m_3^{} \Delta_{31}^{} \left(U_{e 2}^{}
U_{\tau 2}^{} U_{\mu 3}^{*} U_{\tau 3}^{*} + U_{e 3}^{}
U_{\tau 3}^{} U_{\mu 2}^{*} U_{\tau 2}^{*} \right)
\right\} \;, 
\end{eqnarray}
for $\triangle_{\tau}^{\prime}$.
The above analytical approximations of 
$|U_{\alpha i}^{\prime}|^2$ and 
$U_{\alpha i}^{\prime}U_{\beta i}^{\prime *}$ satisfy
$\sum_i^{} |U_{\alpha i}^{\prime}|^2=\sum_{\alpha}^{} 
|U_{\alpha i}^{\prime}|^2 =1$ and $\sum_i^{} U_{\alpha i}^{\prime}
U_{\beta i}^{\prime *}=0$.
The LUTs $\triangle_{\alpha}^{\prime}$ of Majorana
neutrinos can be fixed from Eqs. (34)---(36) though the vector sides 
$U_{\alpha i}^{\prime}U_{\alpha j}^{\prime*}$ 
of $\triangle_i^{\prime}$ can not be derived 
in this way, implying that it is impossible to get any 
information on the Majorana phases at 
$\Lambda_{\rm EW}^{}$. However, we can calculate $|U_{\alpha 
i}^{\prime}U_{\alpha j}^{\prime*}|^2$ 
from Eqs. (31)---(33) and fix the shapes of 
$\triangle_i^{\prime}$ without their orientations. 
With the help of Eq. (4) and the vector sides in Eqs. (34)---(36),
the Jarlskog invariant ${\cal J}^{\prime}$ at $\Lambda_{\rm EW}^{}$
for Majorana neutrinos can be given by
\begin{flalign}
{\cal J}^{\prime}   \simeq \hspace{0.2cm}& 
{\cal J} - \frac{2 \epsilon}{\Delta_{21}^{}
	\Delta_{31}^{} \Delta_{32}^{} } \left\{{\cal J}\left[ m_1^{2}
\left( m_2^{4} + m_3^{4}\right) \left(|U_{\tau 3}^{}|^2
-|U_{\tau 2}^{}|^2 \right) + m_2^{2}
\left( m_1^{4} + m_3^{4}\right) \left(|U_{\tau 1}^{}|^2
-|U_{\tau 3}^{}|^2 \right) \right.\right.&\nonumber\\
& \left. \left. + m_3^{2}
\left( m_1^{4} + m_2^{4}\right) \left(|U_{\tau 2}^{}|^2
-|U_{\tau 1}^{}|^2 \right) \right] 
+  m_1^{} m_2^{}\Delta_{31}^{} \Delta_{32}^{} \left[\left(
|U_{\tau 2}^{}|^2 - |U_{\tau 1}^{}|^2\right) 
{\mathbb I}_{e \tau}^{12} \right.\right.&\nonumber\\
&\left.\left.
- \left(|U_{e 2}^{}|^2 - |U_{e 1}^{}|^2\right) 
{\mathbb I}_{\tau \tau}^{12}\right]+  m_1^{} m_3^{}
\Delta_{21}^{} \Delta_{32}^{} \left[\left(
|U_{\tau 1}^{}|^2 - |U_{\tau 3}^{}|^2\right) 
{\mathbb I}_{e \tau}^{13} - \left(
|U_{e 1}^{}|^2 - |U_{e 3}^{}|^2\right) 
{\mathbb I}_{\tau \tau}^{13}\right]\right.&\nonumber\\
&\left.
+  m_2^{} m_3^{}
\Delta_{21}^{} \Delta_{31}^{} \left[\left(
|U_{\tau 3}^{}|^2 - |U_{\tau 2}^{}|^2\right) 
{\mathbb I}_{e \tau}^{23} - \left(
|U_{e 3}^{}|^2 - |U_{e 2}^{}|^2\right) {\mathbb I}_{\tau \tau}^{23}\right]
\right\}&
\end{flalign}
where ${\mathbb I}_{\alpha\beta}^{ij}$ denote the imaginary
parts of $U_{\alpha i}^{}U_{\alpha j}^{*}U_{\beta i}^{}U_{\beta j}^{*}$.

The nine inner angles of LUTs 
at $\Lambda_{\rm EW}^{}$ running from $\Lambda_{\rm H}^{}$ 
can be derived from Eqs. (34)---(36) and Eq. (21), and expressed
as:
\begin{eqnarray}
\cot \phi_{e1}^{\prime} \hspace{-0.17cm} & \simeq &\hspace{-0.17cm}
\cot \phi_{e1}^{} + \frac{2 \eta_{\phi}^{} \epsilon 
}{{\cal J} \Delta_{21}^{} \Delta_{31}^{} \Delta_{32}^{} }
\left\{
|U_{\tau 2}^{}|^2 |U_{\tau 3}^{}|^2 \left(m_2^{2} \Delta_{31}^{2}
|U_{\mu 3}^{}|^2 -  m_3^{2} \Delta_{21}^{2}
|U_{\mu 2}^{}|^2\right) \right.\nonumber\\  
&& \left. \hspace{-0.17cm} + m_1^{} m_2^{} \Delta_{31} \Delta_{32}
\left[\left(|U_{\tau 1}^{}|^2 + |U_{\tau 2}^{}|^2 \right)
{\mathbb R}_{\mu\tau}^{12} + \left(|U_{\mu 1}^{}|^2 + |U_{\mu 2}^{}|^2 
\right) {\mathbb R}_{\tau\tau}^{12} \right]\right.\nonumber\\  
&& \left. \hspace{-0.17cm} + m_1^{} m_3^{} \Delta_{21} \Delta_{32}
\left[\left(|U_{\tau 1}^{}|^2 + |U_{\tau 3}^{}|^2 \right)
{\mathbb R}_{\mu\tau}^{13} + \left(|U_{\mu 1}^{}|^2 + |U_{\mu 3}^{}|^2 
\right) {\mathbb R}_{\tau\tau}^{13}\right]\right.\nonumber\\  
&& \left. \hspace{-0.17cm} + m_2^{} m_3^{} \Delta_{21} \Delta_{31}
\left[\left(|U_{\tau 3}^{}|^2 - |U_{\tau 2}^{}|^2 \right)
{\mathbb R}_{\mu\tau}^{23} + \left(|U_{\mu 3}^{}|^2 - |U_{\mu 2}^{}|^2 
\right) {\mathbb R}_{\tau\tau}^{23}\right]  \right.\nonumber\\
&& \left. \hspace{-0.17cm} + \cot \phi_{e1}^{} \left[
m_1^{} m_2^{} \Delta_{31} \Delta_{32}
\left(\left(|U_{\tau 1}^{}|^2 - |U_{\tau 2}^{}|^2 \right)
{\mathbb I}_{\mu\tau}^{12} - \left(|U_{\mu 1}^{}|^2 - |U_{\mu 2}^{}|^2 
\right) {\mathbb I}_{\tau\tau}^{12} \right) \right.\right. \nonumber\\
&& \left. \left.\hspace{-0.17cm}+ m_1^{} m_3^{} \Delta_{21} \Delta_{32}
\left(\left(|U_{\tau 3}^{}|^2 - |U_{\tau 1}^{}|^2 \right)
{\mathbb I}_{\mu\tau}^{13} - \left(|U_{\mu 3}^{}|^2 - |U_{\mu 1}^{}|^2 
\right) {\mathbb I}_{\tau\tau}^{13} \right) \right.\right. \nonumber\\
&& \left. \left.\hspace{-0.17cm}+ m_2^{} m_3^{} \Delta_{21} \Delta_{31}
\left(\left(|U_{\tau 2}^{}|^2 - |U_{\tau 3}^{}|^2 \right)
{\mathbb I}_{\mu\tau}^{23} - \left(|U_{\mu 2}^{}|^2 - |U_{\mu 3}^{}|^2 
\right) {\mathbb I}_{\tau\tau}^{23} \right) 
\right]\right\} \;, \nonumber
\end{eqnarray}
\vspace{-0.9cm}
\begin{eqnarray}
\cot \phi_{e2}^{\prime} \hspace{-0.17cm} & \simeq &\hspace{-0.17cm}
\cot \phi_{e2}^{} + \frac{2 \eta_{\phi}^{} \epsilon}{{\cal J} 
	\Delta_{21}^{} \Delta_{31}^{} \Delta_{32}^{} }
\left\{|U_{\tau 1}^{}|^2 |U_{\tau 3}^{}|^2 \left(m_3^{2} 
\Delta_{21}^{2}|U_{\mu 1}^{}|^2 -  m_1^{2} \Delta_{32}^{2}
|U_{\mu 3}^{}|^2\right) \right.\nonumber\\  
&& \left. \hspace{-0.17cm} - m_1^{} m_2^{} \Delta_{31} \Delta_{32}
\left[\left(|U_{\tau 1}^{}|^2 + |U_{\tau 2}^{}|^2 \right)
{\mathbb R}_{\mu\tau}^{12} + \left(|U_{\mu 1}^{}|^2 + |U_{\mu 2}^{}|^2 
\right) {\mathbb R}_{\tau\tau}^{12} \right]\right.\nonumber\\  
&& \left. \hspace{-0.17cm} + m_1^{} m_3^{} \Delta_{21} \Delta_{32}
\left[\left(|U_{\tau 3}^{}|^2 - |U_{\tau 1}^{}|^2 \right)
{\mathbb R}_{\mu\tau}^{13} + \left(|U_{\mu 3}^{}|^2 - |U_{\mu 1}^{}|^2 
\right) {\mathbb R}_{\tau\tau}^{13}\right]\right.\nonumber\\  
&& \left. \hspace{-0.17cm} + m_2^{} m_3^{} \Delta_{21} \Delta_{31}
\left[\left(|U_{\tau 2}^{}|^2 + |U_{\tau 3}^{}|^2 \right)
{\mathbb R}_{\mu\tau}^{23} + \left(|U_{\mu 2}^{}|^2 + |U_{\mu 3}^{}|^2 
\right) {\mathbb R}_{\tau\tau}^{23}\right]  \right.\nonumber\\
&& \left. \hspace{-0.17cm} + \cot \phi_{e2}^{} \left[
m_1^{} m_2^{} \Delta_{31} \Delta_{32}
\left(\left(|U_{\tau 1}^{}|^2 - |U_{\tau 2}^{}|^2 \right)
{\mathbb I}_{\mu\tau}^{12} - \left(|U_{\mu 1}^{}|^2 - |U_{\mu 2}^{}|^2 
\right) {\mathbb I}_{\tau\tau}^{12} \right) \right.\right. \nonumber\\
&& \left. \left.\hspace{-0.17cm}+ m_1^{} m_3^{} \Delta_{21} \Delta_{32}
\left(\left(|U_{\tau 3}^{}|^2 - |U_{\tau 1}^{}|^2 \right)
{\mathbb I}_{\mu\tau}^{13} - \left(|U_{\mu 3}^{}|^2 - |U_{\mu 1}^{}|^2 
\right) {\mathbb I}_{\tau\tau}^{13} \right) \right.\right. \nonumber\\
&& \left. \left.\hspace{-0.17cm}+ m_2^{} m_3^{} \Delta_{21} \Delta_{31}
\left(\left(|U_{\tau 2}^{}|^2 - |U_{\tau 3}^{}|^2 \right)
{\mathbb I}_{\mu\tau}^{23} - \left(|U_{\mu 2}^{}|^2 - |U_{\mu 3}^{}|^2 
\right) {\mathbb I}_{\tau\tau}^{23} \right) 
\right]\right\} \;, \nonumber
\end{eqnarray}
\vspace{-0.9cm}
\begin{eqnarray}
\cot \phi_{e3}^{\prime} \hspace{-0.17cm} & \simeq &\hspace{-0.17cm}
\cot \phi_{e3}^{} + \frac{2 \eta_{\phi}^{} \epsilon}{{\cal J} 
	\Delta_{21}^{} \Delta_{31}^{} \Delta_{32}^{} }
\left\{|U_{\tau 1}^{}|^2 |U_{\tau 2}^{}|^2 \left(m_1^{2} 
\Delta_{32}^{2}|U_{\mu 2}^{}|^2 -  m_2^{2} \Delta_{31}^{2}
|U_{\mu 1}^{}|^2\right) \right.\nonumber\\  
&& \left. \hspace{-0.17cm} + m_1^{} m_2^{} \Delta_{31} \Delta_{32}
\left[\left(|U_{\tau 2}^{}|^2 - |U_{\tau 1}^{}|^2 \right)
{\mathbb R}_{\mu\tau}^{12} + \left(|U_{\mu 2}^{}|^2 - |U_{\mu 1}^{}|^2 
\right) {\mathbb R}_{\tau\tau}^{12} \right]\right.\nonumber\\  
&& \left. \hspace{-0.17cm} - m_1^{} m_3^{} \Delta_{21} \Delta_{32}
\left[\left(|U_{\tau 1}^{}|^2 + |U_{\tau 3}^{}|^2 \right)
{\mathbb R}_{\mu\tau}^{13} + \left(|U_{\mu 1}^{}|^2 + |U_{\mu 3}^{}|^2 
\right) {\mathbb R}_{\tau\tau}^{13}\right]\right.\nonumber\\  
&& \left. \hspace{-0.17cm} - m_2^{} m_3^{} \Delta_{21} \Delta_{31}
\left[\left(|U_{\tau 2}^{}|^2 + |U_{\tau 3}^{}|^2 \right)
{\mathbb R}_{\mu\tau}^{23} + \left(|U_{\mu 2}^{}|^2 + |U_{\mu 3}^{}|^2 
\right) {\mathbb R}_{\tau\tau}^{23}\right]  \right.\nonumber\\
&& \left. \hspace{-0.17cm} + \cot \phi_{e3}^{} \left[
m_1^{} m_2^{} \Delta_{31} \Delta_{32}
\left(\left(|U_{\tau 1}^{}|^2 - |U_{\tau 2}^{}|^2 \right)
{\mathbb I}_{\mu\tau}^{12} - \left(|U_{\mu 1}^{}|^2 - |U_{\mu 2}^{}|^2 
\right) {\mathbb I}_{\tau\tau}^{12} \right) \right.\right. \nonumber\\
&& \left. \left.\hspace{-0.17cm}+ m_1^{} m_3^{} \Delta_{21} \Delta_{32}
\left(\left(|U_{\tau 3}^{}|^2 - |U_{\tau 1}^{}|^2 \right)
{\mathbb I}_{\mu\tau}^{13} - \left(|U_{\mu 3}^{}|^2 - |U_{\mu 1}^{}|^2 
\right) {\mathbb I}_{\tau\tau}^{13} \right) \right.\right. \nonumber\\
&& \left. \left.\hspace{-0.17cm}+ m_2^{} m_3^{} \Delta_{21} \Delta_{31}
\left(\left(|U_{\tau 2}^{}|^2 - |U_{\tau 3}^{}|^2 \right)
{\mathbb I}_{\mu\tau}^{23} - \left(|U_{\mu 2}^{}|^2 - |U_{\mu 3}^{}|^2 
\right) {\mathbb I}_{\tau\tau}^{23} \right) 
\right]\right\} \;; 
\end{eqnarray}
and
\begin{eqnarray}
\cot \phi_{\mu 1}^{\prime} \hspace{-0.17cm} & \simeq &\hspace{-0.17cm}
\cot \phi_{\mu 1}^{} + \frac{2 \eta_{\phi}^{} \epsilon }{{\cal J} 
	\Delta_{21}^{} \Delta_{31}^{} \Delta_{32}^{} }
\left\{
|U_{\tau 2}^{}|^2 |U_{\tau 3}^{}|^2 \left(m_2^{2} \Delta_{31}^{2}
|U_{e 3}^{}|^2 -  m_3^{2} \Delta_{21}^{2}
|U_{e 2}^{}|^2\right) \right.\nonumber\\  
&& \left. \hspace{-0.17cm} + m_1^{} m_2^{} \Delta_{31} \Delta_{32}
\left[\left(|U_{\tau 1}^{}|^2 + |U_{\tau 2}^{}|^2 \right)
{\mathbb R}_{e\tau}^{12} + \left(|U_{e 1}^{}|^2 + |U_{e 2}^{}|^2 
\right) {\mathbb R}_{\tau\tau}^{12} \right]\right.\nonumber\\  
&& \left. \hspace{-0.17cm} + m_1^{} m_3^{} \Delta_{21} \Delta_{32}
\left[\left(|U_{\tau 1}^{}|^2 + |U_{\tau 3}^{}|^2 \right)
{\mathbb R}_{e\tau}^{13} + \left(|U_{e 1}^{}|^2 + |U_{e 3}^{}|^2 
\right) {\mathbb R}_{\tau\tau}^{13}\right]\right.\nonumber\\  
&& \left. \hspace{-0.17cm} + m_2^{} m_3^{} \Delta_{21} \Delta_{31}
\left[\left(|U_{\tau 3}^{}|^2 - |U_{\tau 2}^{}|^2 \right)
{\mathbb R}_{e\tau}^{23} + \left(|U_{e 3}^{}|^2 - |U_{e 2}^{}|^2 
\right) {\mathbb R}_{\tau\tau}^{23}\right]  \right.\nonumber\\
&& \left. \hspace{-0.17cm} - \cot \phi_{\mu 1}^{} \left[
m_1^{} m_2^{} \Delta_{31} \Delta_{32}
\left(\left(|U_{\tau 1}^{}|^2 - |U_{\tau 2}^{}|^2 \right)
{\mathbb I}_{e\tau}^{12} - \left(|U_{e 1}^{}|^2 - |U_{e 2}^{}|^2 
\right) {\mathbb I}_{\tau\tau}^{12} \right) \right.\right. \nonumber\\
&& \left. \left.\hspace{-0.17cm}+ m_1^{} m_3^{} \Delta_{21} \Delta_{32}
\left(\left(|U_{\tau 3}^{}|^2 - |U_{\tau 1}^{}|^2 \right)
{\mathbb I}_{e\tau}^{13} - \left(|U_{e 3}^{}|^2 - |U_{e 1}^{}|^2 
\right) {\mathbb I}_{\tau\tau}^{13} \right) \right.\right. \nonumber\\
&& \left. \left.\hspace{-0.17cm}+ m_2^{} m_3^{} \Delta_{21} \Delta_{31}
\left(\left(|U_{\tau 2}^{}|^2 - |U_{\tau 3}^{}|^2 \right)
{\mathbb I}_{e\tau}^{23} - \left(|U_{e 2}^{}|^2 - |U_{e 3}^{}|^2 
\right) {\mathbb I}_{\tau\tau}^{23} \right) 
\right]\right\} \;, \nonumber
\end{eqnarray}
\vspace{-0.9cm}
\begin{eqnarray}
\cot \phi_{\mu 2}^{\prime} \hspace{-0.17cm} & \simeq &\hspace{-0.17cm}
\cot \phi_{\mu 2}^{} +\frac{2 \eta_{\phi}^{} \epsilon}{{\cal J} 
	\Delta_{21}^{} \Delta_{31}^{} \Delta_{32}^{} }
\left\{
|U_{\tau 1}^{}|^2 |U_{\tau 3}^{}|^2 \left(m_3^{2} \Delta_{21}^{2}
|U_{e 1}^{}|^2 -  m_1^{2} \Delta_{32}^{2}
|U_{e 3}^{}|^2\right) \right.\nonumber\\  
&& \left. \hspace{-0.17cm} - m_1^{} m_2^{} \Delta_{31} \Delta_{32}
\left[\left(|U_{\tau 1}^{}|^2 + |U_{\tau 2}^{}|^2 \right)
{\mathbb R}_{e\tau}^{12} + \left(|U_{e 1}^{}|^2 + |U_{e 2}^{}|^2 
\right) {\mathbb R}_{\tau\tau}^{12} \right]\right.\nonumber\\  
&& \left. \hspace{-0.17cm} + m_1^{} m_3^{} \Delta_{21} \Delta_{32}
\left[\left(|U_{\tau 3}^{}|^2 - |U_{\tau 1}^{}|^2 \right)
{\mathbb R}_{e\tau}^{13} + \left(|U_{e 3}^{}|^2 - |U_{e 1}^{}|^2 
\right) {\mathbb R}_{\tau\tau}^{13}\right]\right.\nonumber\\  
&& \left. \hspace{-0.17cm} + m_2^{} m_3^{} \Delta_{21} \Delta_{31}
\left[\left(|U_{\tau 2}^{}|^2 + |U_{\tau 3}^{}|^2 \right)
{\mathbb R}_{e\tau}^{23} + \left(|U_{e 2}^{}|^2 + |U_{e 3}^{}|^2 
\right) {\mathbb R}_{\tau\tau}^{23}\right]  \right.\nonumber\\
&& \left. \hspace{-0.17cm} - \cot \phi_{\mu 2}^{} \left[
m_1^{} m_2^{} \Delta_{31} \Delta_{32}
\left(\left(|U_{\tau 1}^{}|^2 - |U_{\tau 2}^{}|^2 \right)
{\mathbb I}_{e\tau}^{12} - \left(|U_{e 1}^{}|^2 - |U_{e 2}^{}|^2 
\right) {\mathbb I}_{\tau\tau}^{12} \right) \right.\right. \nonumber\\
&& \left. \left.\hspace{-0.17cm}+ m_1^{} m_3^{} \Delta_{21} \Delta_{32}
\left(\left(|U_{\tau 3}^{}|^2 - |U_{\tau 1}^{}|^2 \right)
{\mathbb I}_{e\tau}^{13} - \left(|U_{e 3}^{}|^2 - |U_{e 1}^{}|^2 
\right) {\mathbb I}_{\tau\tau}^{13} \right) \right.\right. \nonumber\\
&& \left. \left.\hspace{-0.17cm}+ m_2^{} m_3^{} \Delta_{21} \Delta_{31}
\left(\left(|U_{\tau 2}^{}|^2 - |U_{\tau 3}^{}|^2 \right)
{\mathbb I}_{e\tau}^{23} - \left(|U_{e 2}^{}|^2 - |U_{e 3}^{}|^2 
\right) {\mathbb I}_{\tau\tau}^{23} \right) 
\right]\right\} \;, \nonumber
\end{eqnarray}
\vspace{-0.9cm}
\begin{eqnarray}
\cot \phi_{\mu 3}^{\prime} \hspace{-0.17cm} & \simeq &\hspace{-0.17cm}
\cot \phi_{\mu 3}^{} + \frac{2 \eta_{\phi}^{}\epsilon}{{\cal J} 
	\Delta_{21}^{} \Delta_{31}^{} \Delta_{32}^{} } \left\{
|U_{\tau 1}^{}|^2 |U_{\tau 2}^{}|^2 \left(m_1^{2} \Delta_{32}^{2}
|U_{e 2}^{}|^2 -  m_2^{2} \Delta_{31}^{2}
|U_{e 1}^{}|^2\right) \right.\nonumber\\  
&& \left. \hspace{-0.17cm} + m_1^{} m_2^{} \Delta_{31} \Delta_{32}
\left[\left(|U_{\tau 2}^{}|^2 - |U_{\tau 1}^{}|^2 \right)
{\mathbb R}_{e\tau}^{12} + \left(|U_{e 2}^{}|^2 - |U_{e 1}^{}|^2 
\right) {\mathbb R}_{\tau\tau}^{12} \right]\right.\nonumber\\  
&& \left. \hspace{-0.17cm} - m_1^{} m_3^{} \Delta_{21} \Delta_{32}
\left[\left(|U_{\tau 1}^{}|^2 + |U_{\tau 3}^{}|^2 \right)
{\mathbb R}_{e\tau}^{13} + \left(|U_{e 1}^{}|^2 + |U_{e 3}^{}|^2 
\right) {\mathbb R}_{\tau\tau}^{13}\right]\right.\nonumber\\  
&& \left. \hspace{-0.17cm} - m_2^{} m_3^{} \Delta_{21} \Delta_{31}
\left[\left(|U_{\tau 2}^{}|^2 + |U_{\tau 3}^{}|^2 \right)
{\mathbb R}_{e\tau}^{23} + \left(|U_{e 2}^{}|^2 + |U_{e 3}^{}|^2 
\right) {\mathbb R}_{\tau\tau}^{23}\right]  \right.\nonumber\\
&& \left. \hspace{-0.17cm} - \cot \phi_{\mu 3}^{} \left[
m_1^{} m_2^{} \Delta_{31} \Delta_{32}
\left(\left(|U_{\tau 1}^{}|^2 - |U_{\tau 2}^{}|^2 \right)
{\mathbb I}_{e\tau}^{12} - \left(|U_{e 1}^{}|^2 - |U_{e 2}^{}|^2 
\right) {\mathbb I}_{\tau\tau}^{12} \right) \right.\right. \nonumber\\
&& \left. \left.\hspace{-0.17cm}+ m_1^{} m_3^{} \Delta_{21} \Delta_{32}
\left(\left(|U_{\tau 3}^{}|^2 - |U_{\tau 1}^{}|^2 \right)
{\mathbb I}_{e\tau}^{13} - \left(|U_{e 3}^{}|^2 - |U_{e 1}^{}|^2 
\right) {\mathbb I}_{\tau\tau}^{13} \right) \right.\right. \nonumber\\
&& \left. \left.\hspace{-0.17cm}+ m_2^{} m_3^{} \Delta_{21} \Delta_{31}
\left(\left(|U_{\tau 2}^{}|^2 - |U_{\tau 3}^{}|^2 \right)
{\mathbb I}_{e\tau}^{23} - \left(|U_{e 2}^{}|^2 - |U_{e 3}^{}|^2 
\right) {\mathbb I}_{\tau\tau}^{23} \right) 
\right]\right\} \;;
\end{eqnarray}
and
\begin{eqnarray}
\cot \phi_{\tau 1}^{\prime} \hspace{-0.17cm} & \simeq &\hspace{-0.17cm}
\cot \phi_{\tau 1}^{} + \frac{2 \eta_{\phi}^{}\epsilon}{{\cal J} 
	\Delta_{21}^{} \Delta_{31}^{} \Delta_{32}^{} }
\left\{  \right. \nonumber\\
&& \left. m_3^{2} \Delta_{21}^{2}
|U_{e 2}^{}|^2 |U_{\mu 2}^{}|^2 \left(1 - |U_{\tau 3}^{}|^2\right) 
- m_2^{2} \Delta_{31}^{2}
|U_{e 3}^{}|^2 |U_{\mu 3}^{}|^2 \left(1 - |U_{\tau 2}^{}|^2\right) 
\right.\nonumber\\  
&& \left. \hspace{-0.17cm} + m_1^{} m_2^{} \Delta_{31} \Delta_{32}
\left[\left(|U_{\mu 1}^{}|^2 + |U_{\mu 2}^{}|^2 \right)
{\mathbb R}_{e\tau}^{12} + \left(|U_{e 1}^{}|^2 + |U_{e 2}^{}|^2 
\right) {\mathbb R}_{\mu\tau}^{12} \right]\right.\nonumber\\  
&& \left. \hspace{-0.17cm} + m_1^{} m_3^{} \Delta_{21} \Delta_{32}
\left[\left(|U_{\mu 1}^{}|^2 + |U_{\mu 3}^{}|^2 \right)
{\mathbb R}_{e\tau}^{13} + \left(|U_{e 1}^{}|^2 + |U_{e 3}^{}|^2 
\right) {\mathbb R}_{\mu\tau}^{13}\right]\right.\nonumber\\  
&& \left. \hspace{-0.17cm} + m_2^{} m_3^{} \Delta_{21} \Delta_{31}
\left[\left(|U_{\mu 3}^{}|^2 - |U_{\mu 2}^{}|^2 \right)
{\mathbb R}_{e\tau}^{23} + \left(|U_{e 3}^{}|^2 - |U_{e 2}^{}|^2 
\right) {\mathbb R}_{\mu\tau}^{23}\right]  \right.\nonumber\\
&& \left. \hspace{-0.17cm} + \cot \phi_{\tau 1}^{} \left[
m_1^{} m_2^{} \Delta_{31} \Delta_{32}
\left(\left(|U_{\mu 1}^{}|^2 - |U_{\mu 2}^{}|^2 \right)
{\mathbb I}_{e\tau}^{12} - \left(|U_{e 1}^{}|^2 - |U_{e 2}^{}|^2 
\right) {\mathbb I}_{\mu\tau}^{12} \right) \right.\right. \nonumber\\
&& \left. \left.\hspace{-0.17cm}+ m_1^{} m_3^{} \Delta_{21} \Delta_{32}
\left(\left(|U_{\mu 3}^{}|^2 - |U_{\mu 1}^{}|^2 \right)
{\mathbb I}_{e\tau}^{13} - \left(|U_{e 3}^{}|^2 - |U_{e 1}^{}|^2 
\right) {\mathbb I}_{\mu\tau}^{13} \right) \right.\right. \nonumber\\
&& \left. \left.\hspace{-0.17cm}+ m_2^{} m_3^{} \Delta_{21} \Delta_{31}
\left(\left(|U_{\mu 2}^{}|^2 - |U_{\mu 3}^{}|^2 \right)
{\mathbb I}_{e\tau}^{23} - \left(|U_{e 2}^{}|^2 - |U_{e 3}^{}|^2 
\right) {\mathbb I}_{\mu\tau}^{23} \right) 
\right]\right\} \;, \nonumber
\end{eqnarray}
\vspace{-0.9cm}
\begin{eqnarray}
\cot \phi_{\tau 2}^{\prime} \hspace{-0.17cm} & \simeq &\hspace{-0.17cm}
\cot \phi_{\tau 2}^{} + \frac{2 \eta_{\phi}^{} \epsilon}{{\cal J} \Delta_{21}^{} 
	\Delta_{31}^{} \Delta_{32}^{} }
\left\{ \right.\nonumber\\
&& \left.m_1^{2} \Delta_{32}^{2}
|U_{e 3}^{}|^2 |U_{\mu 3}^{}|^2 \left(1 - |U_{\tau 1}^{}|^2\right) 
- m_3^{2} \Delta_{21}^{2}
|U_{e 1}^{}|^2 |U_{\mu 1}^{}|^2 \left(1 - |U_{\tau 3}^{}|^2\right) 
\right.\nonumber\\  
&& \left. \hspace{-0.17cm} - m_1^{} m_2^{} \Delta_{31} \Delta_{32}
\left[\left(|U_{\mu 1}^{}|^2 + |U_{\mu 2}^{}|^2 \right)
{\mathbb R}_{e\tau}^{12} + \left(|U_{e 1}^{}|^2 + |U_{e 2}^{}|^2 
\right) {\mathbb R}_{\mu\tau}^{12} \right]\right.\nonumber\\  
&& \left. \hspace{-0.17cm} + m_1^{} m_3^{} \Delta_{21} \Delta_{32}
\left[\left(|U_{\mu 3}^{}|^2 - |U_{\mu 1}^{}|^2 \right)
{\mathbb R}_{e\tau}^{13} + \left(|U_{e 3}^{}|^2 - |U_{e 1}^{}|^2 
\right) {\mathbb R}_{\mu\tau}^{13}\right]\right.\nonumber\\  
&& \left. \hspace{-0.17cm} + m_2^{} m_3^{} \Delta_{21} \Delta_{31}
\left[\left(|U_{\mu 2}^{}|^2 + |U_{\mu 3}^{}|^2 \right)
{\mathbb R}_{e\tau}^{23} + \left(|U_{e 2}^{}|^2 + |U_{e 3}^{}|^2 
\right) {\mathbb R}_{\mu\tau}^{23}\right]  \right.\nonumber\\
&& \left. \hspace{-0.17cm} + \cot \phi_{\tau 2}^{} \left[
m_1^{} m_2^{} \Delta_{31} \Delta_{32}
\left(\left(|U_{\mu 1}^{}|^2 - |U_{\mu 2}^{}|^2 \right)
{\mathbb I}_{e\tau}^{12} - \left(|U_{e 1}^{}|^2 - |U_{e 2}^{}|^2 
\right) {\mathbb I}_{\mu\tau}^{12} \right) \right.\right. \nonumber\\
&& \left. \left.\hspace{-0.17cm}+ m_1^{} m_3^{} \Delta_{21} \Delta_{32}
\left(\left(|U_{\mu 3}^{}|^2 - |U_{\mu 1}^{}|^2 \right)
{\mathbb I}_{e\tau}^{13} - \left(|U_{e 3}^{}|^2 - |U_{e 1}^{}|^2 
\right) {\mathbb I}_{\mu\tau}^{13} \right) \right.\right. \nonumber\\
&& \left. \left.\hspace{-0.17cm}+ m_2^{} m_3^{} \Delta_{21} \Delta_{31}
\left(\left(|U_{\mu 2}^{}|^2 - |U_{\mu 3}^{}|^2 \right)
{\mathbb I}_{e\tau}^{23} - \left(|U_{e 2}^{}|^2 - |U_{e 3}^{}|^2 
\right) {\mathbb I}_{\mu\tau}^{23} \right) 
\right]\right\} \;, \nonumber
\end{eqnarray}
\vspace{-0.9cm}
\begin{eqnarray}
\cot \phi_{\tau 3}^{\prime} \hspace{-0.17cm} & \simeq &\hspace{-0.17cm}
\cot \phi_{\tau 3}^{} + \frac{2 \eta_{\phi}^{}\epsilon}
{{\cal J} \Delta_{21}^{} \Delta_{31}^{} \Delta_{32}^{} }
\left\{ \right.\nonumber\\
&&\left. m_2^{2} \Delta_{31}^{2}
|U_{e 1}^{}|^2 |U_{\mu 1}^{}|^2 \left(1 - |U_{\tau 2}^{}|^2\right) 
- m_1^{2} \Delta_{32}^{2}
|U_{e 2}^{}|^2 |U_{\mu 2}^{}|^2 \left(1 - |U_{\tau 1}^{}|^2\right) 
\right.\nonumber\\  
&& \left. \hspace{-0.17cm} + m_1^{} m_2^{} \Delta_{31} \Delta_{32}
\left[\left(|U_{\mu 2}^{}|^2 - |U_{\mu 1}^{}|^2 \right)
{\mathbb R}_{e\tau}^{12} + \left(|U_{e 2}^{}|^2 - |U_{e 1}^{}|^2 
\right) {\mathbb R}_{\mu\tau}^{12} \right]\right.\nonumber\\  
&& \left. \hspace{-0.17cm} - m_1^{} m_3^{} \Delta_{21} \Delta_{32}
\left[\left(|U_{\mu 1}^{}|^2 + |U_{\mu 3}^{}|^2 \right)
{\mathbb R}_{e\tau}^{13} + \left(|U_{e 1}^{}|^2 + |U_{e 3}^{}|^2 
\right) {\mathbb R}_{\mu\tau}^{13}\right]\right.\nonumber\\  
&& \left. \hspace{-0.17cm} - m_2^{} m_3^{} \Delta_{21} \Delta_{31}
\left[\left(|U_{\mu 2}^{}|^2 + |U_{\mu 3}^{}|^2 \right)
{\mathbb R}_{e\tau}^{23} + \left(|U_{e 2}^{}|^2 + |U_{e 3}^{}|^2 
\right) {\mathbb R}_{\mu\tau}^{23}\right]  \right.\nonumber\\
&& \left. \hspace{-0.17cm} + \cot \phi_{\tau 3}^{} \left[
m_1^{} m_2^{} \Delta_{31} \Delta_{32}
\left(\left(|U_{\mu 1}^{}|^2 - |U_{\mu 2}^{}|^2 \right)
{\mathbb I}_{e\tau}^{12} - \left(|U_{e 1}^{}|^2 - |U_{e 2}^{}|^2 
\right) {\mathbb I}_{\mu\tau}^{12} \right) \right.\right. \nonumber\\
&& \left. \left.\hspace{-0.17cm}+ m_1^{} m_3^{} \Delta_{21} \Delta_{32}
\left(\left(|U_{\mu 3}^{}|^2 - |U_{\mu 1}^{}|^2 \right)
{\mathbb I}_{e\tau}^{13} - \left(|U_{e 3}^{}|^2 - |U_{e 1}^{}|^2 
\right) {\mathbb I}_{\mu\tau}^{13} \right) \right.\right. \nonumber\\
&& \left. \left.\hspace{-0.17cm}+ m_2^{} m_3^{} \Delta_{21} \Delta_{31}
\left(\left(|U_{\mu 2}^{}|^2 - |U_{\mu 3}^{}|^2 \right)
{\mathbb I}_{e\tau}^{23} - \left(|U_{e 2}^{}|^2 - |U_{e 3}^{}|^2 
\right) {\mathbb I}_{\mu\tau}^{23} \right) 
\right]\right\} \;.
\end{eqnarray}

Some discussions about the analytical results 
above for both Dirac and Majorana neutrinos are as follows:
\begin{itemize}
\item 
The approximate expressions of $|U_{e i}^{\prime}|^2$ 
and $U_{\mu i}^{\prime}U_{\tau i}^{\prime*}$ 
are similar to those of $|U_{\mu i}^{\prime}|^2$ 
and $U_{\tau i}^{\prime}U_{e i}^{\prime*}$, 
respectively. 
The analytical results for Majorana neutrinos are not
equivalent to those for Dirac neutrinos even if one
turns off the Majorana phases by setting their values
to be zeros. In both cases, the corrections to
the LUTs depend a lot on the magnitudes of 
the lightest neutrino mass
and the small quantity $\epsilon$. The evolutions of the sides
$U_{\alpha 3}^{\prime} U_{\beta 3}^{\prime *}$ and the inner 
angles $\phi_{\alpha 3}$ are more stable against the RGE running. 
\item 
Different from the Dirac case, 
${\cal J}^{\prime}$ of Majorana neutrinos is in general
nonzero
even assuming ${\cal J}$ at $\Lambda_{\rm H}^{}$ to be zero, 
and vice versa. 
One can conclude from Eq. (37) that there may
 exist leptonic CP violation at 
$\Lambda_{\rm EW}^{}$ unless all the Dirac and Majorana 
phases at the superhigh energy vanish.
This observation is consistent with 
the analysis in Refs. \cite {Luo,Xing:2005fw}.
\item
The direct connections of the LUTs between two energy 
scales, which have been established above, are independent of
the parametrization of $U$ and complementary to the 
differential forms in Ref. \cite{Luo}. 
They can also reproduce the analytical approximations of
neutrino masses, flavor mixing angles and the Dirac CP
phase in other references \cite{Huang,Xing:2017mkx,Zhou:2014sya} 
by taking a specific
parametrization. Note that the accuracy of 
the approximate results above and in section 3 will be very poor 
if the neutrino masses are strongly degenerate, i.e., 
the smallest neutrino mass is big enough. 
Considering the fact that the combination of Planck and
baryon acoustic oscillation (BAO) measurements gives
the limit of the sum
of three light neutrino masses as $\sum_i^{} m_i^{} <
0.12 $ eV at $95\%$ confidence level \cite{Planck},
one can use the analytical approximations to understand
most part of the parameter space.
We plan to explicitly study the case of nearly degenerate
neutrino masses elsewhere.
\end{itemize}

\section{LUTs and RGE-induced $\mu$-$\tau$ reflection symmetry breaking}
The $\mu$-$\tau$ reflection symmetry of the neutrino sector
serving as the minimal discrete flavor symmetry to explain the
lepton flavor mixing and CP violation has been 
extensively studied for both Dirac and Majorana neutrinos 
\cite{Xing:2015fdg}. One of the usual ways is that 
by assuming the $\mu$-$\tau$ reflection symmetry at a superhigh 
energy scale $\Lambda_{\mu\tau}^{}$, we 
confront its RGE-induced breaking effects 
at $\Lambda_{\rm EW}^{}$ 
with current experiment data 
\cite{Xing:2015fdg,Xing:2017mkx,mutauRGE,Nath:2018hjx,Zhou:2014sya}.
This can be connected with the corresponding reformations of the LUTs below.
\subsection{The case of Dirac neutrinos}
If massive neutrinos are the Dirac particles, 
the $\mu$-$\tau$ reflection symmetry
means that the effective Dirac neutrino mass term is invariant 
under the flavor and charge-conjugation transformations below: 
\begin{align}
\nu^{}_{e \rm L} &\leftrightarrow (\nu^{}_{e \rm L})^{\rm c} \;, 
& \nu^{}_{\mu \rm L} &\leftrightarrow (\nu^{}_{\tau \rm L})^{\rm c} \;,
& \nu^{}_{\tau \rm L} &\leftrightarrow (\nu^{}_{\mu \rm L})^{\rm c} \;,
\nonumber\\
N^{}_{e \rm R} &\leftrightarrow (N^{}_{e \rm R})^{\rm c} \;,
& N^{}_{\mu \rm R} &\leftrightarrow (N^{}_{\tau \rm R})^{\rm c} \;,
& N^{}_{\tau \rm R} &\leftrightarrow (N^{}_{\mu \rm R})^{\rm c} \;,
\end{align}
where $\nu_{\alpha {\rm L}}$ and $N_{\alpha {\rm R}}$
for  $\alpha = e, \mu,\tau$ denote the left-handed 
and right-handed neutrino fields, respectively. 
This results in the constraint
conditions of $(H_{\nu}^{})_{e\mu}^{}=(H_{\nu}^{})_{e\tau}^{*}$
and $(H_{\nu}^{})_{\mu\mu}^{}=(H_{\nu}^{})_{\tau\tau}^{}$
with $(H_{\nu}^{})_{\alpha\beta}^{}=\sum_{i} m_i^2
U_{\alpha i}^{} U_{\beta i}^{*}$ being defined in 
subsection 2.1. To be specific, we have 
$U_{e i}^{} U_{\mu i}^{*}=U_{e i}^{*} U_{\tau i}^{}$
and $|U_{\mu i}^{}|=|U_{\tau i}^{}|$ for 
$i=1,2,3$, which can also be expressed as $U_{ei}^{}=\eta_i U_{ei}^{*}$ 
and $U_{\mu i}^{}=\eta_i U_{\tau i}^{*}$ with $\eta_i=\pm 1$.
There are eight choices of $(\eta_1^{},\eta_2^{},\eta_3^{})$
while all of them are identical with one another through
blackefining the relevant phases of charged lepton and Dirac
neutrino fields. Given the $\mu$-$\tau$ reflection symmetry of
Dirac neutrinos at a superhigh energy scale 
$\Lambda_{\mu\tau}^{}$,
we have $|U_{\mu i}^{}|=|U_{\tau i}^{}|$. Hence
the corresponding $\triangle_i^{}$
are isosceles triangles, each with two equal sides 
$|U_{\mu j}^{} U_{\mu k}^{*}|=
|U_{\tau j}^{} U_{\tau k}^{*}|$;
and the two LUTs $\triangle_{\mu}^{}$
and $\triangle_{\tau}^{}$ are congruent 
with each other with three pairs of equal sides $|U_{\tau i}^{} U_{e i}^{*}|=
|U_{e i}^{} U_{\mu i}^{*}|$.
The deviations of the LUTs at $\Lambda_{\rm EW}^{}$ from 
these special shapes at $\Lambda_{\mu\tau}^{}$ due to the
RGE running
can demonstrate the RGE-induced $\mu$-$\tau$ reflection symmetry 
breaking intuitively. Let us define 
\begin{eqnarray}
{\cal S}_{\triangle_1^{}}^{\mu\tau} \hspace{-0.17cm} & \equiv
&\hspace{-0.17cm} |U_{\mu 2}^{\prime} U_{\mu 3}^{\prime*}|^2 
- |U_{\tau 2}^{\prime} U_{\tau 3}^{\prime*}|^2 \;,\nonumber \\
{\cal S}_{\triangle_2^{}}^{\mu\tau} \hspace{-0.17cm} & \equiv
&\hspace{-0.17cm} |U_{\mu 3}^{\prime} U_{\mu 1}^{\prime*}|^2 
- |U_{\tau 3}^{\prime} U_{\tau 1}^{\prime*}|^2 \;,\nonumber \\
{\cal S}_{\triangle_3^{}}^{\mu\tau} \hspace{-0.17cm} & \equiv
&\hspace{-0.17cm} |U_{\mu 1}^{\prime} U_{\mu 2}^{\prime*}|^2 
- |U_{\tau 1}^{\prime} U_{\tau 2}^{\prime*}|^2 \;, 
\end{eqnarray}
to describe the deviations of $\Delta_{i}^{\prime}$
from their $\mu$-$\tau$ reflection symmetry limits, and
\begin{eqnarray}
{\cal S}_{\triangle_{\mu\tau}^{}}^{1} \hspace{-0.17cm} & \equiv
&\hspace{-0.17cm} |U_{\tau 1}^{\prime} U_{e 1}^{\prime*}|^2 
- |U_{e 1}^{\prime} U_{\mu 1}^{\prime*}|^2 \;,\nonumber \\
{\cal S}_{\triangle_{\mu\tau}^{}}^{2} \hspace{-0.17cm} & \equiv
&\hspace{-0.17cm} |U_{\tau 2}^{\prime} U_{e 2}^{\prime*}|^2 
- |U_{e 2}^{\prime} U_{\mu 2}^{\prime*}|^2 \;,\nonumber \\
{\cal S}_{\triangle_{\mu\tau}^{}}^{3} \hspace{-0.17cm} & \equiv
&\hspace{-0.17cm} |U_{\tau 3}^{\prime} U_{e 3}^{\prime*}|^2 
- |U_{e 3}^{\prime} U_{\mu 3}^{\prime*}|^2  \;,
\end{eqnarray}
to show how the LUTs $\Delta_{\mu}^{\prime}$
and $\Delta_{\tau}^{\prime}$ can be reformed
as compablack with their $\mu$-$\tau$ reflection symmetry limits.
With the help of $|U_{\mu i}^{}|=|U_{\tau i}^{}|$ 
together with Eqs. (14)---(16),
the analytical approximations of the six
asymmetries in Eqs. (42) and (43) can be expressed as:
\begin{eqnarray}
{\cal S}_{\triangle_1^{}}^{\mu\tau} \hspace{-0.17cm} & \simeq 
&\hspace{-0.17cm}
\frac{\epsilon}{\Delta_{21}^{} \Delta_{31}^{} \Delta_{32}^{}}
\left[m_2^{2} \Delta_{31}^{2} |U_{e3}^{}|^2 \left( 1 - |U_{e 3}^{}|^2 \right)
- m_3^{2} \Delta_{21}^{2} |U_{e2}^{}|^2 \left( 1 - |U_{e 2}^{}|^2 \right)
\right]\;,\nonumber\\
{\cal S}_{\triangle_2^{}}^{\mu\tau} \hspace{-0.17cm} & \simeq 
&\hspace{-0.17cm}
\frac{\epsilon}{\Delta_{21}^{} \Delta_{31}^{} \Delta_{32}^{}}
\left[m_3^{2} \Delta_{21}^{2} |U_{e1}^{}|^2 \left( 1 - |U_{e 1}^{}|^2 \right) 
- m_1^{2} \Delta_{32}^{2} |U_{e3}^{}|^2 \left( 1 - |U_{e 3}^{}|^2 \right)
\right]\;,\nonumber\\
{\cal S}_{\triangle_3^{}}^{\mu\tau} \hspace{-0.17cm} & \simeq 
&\hspace{-0.17cm}
\frac{\epsilon}{\Delta_{21}^{} \Delta_{31}^{} \Delta_{32}^{}}
\left[m_1^{2} \Delta_{32}^{2} |U_{e2}^{}|^2 \left( 1 - |U_{e 2}^{}|^2 \right)
- m_2^{2} \Delta_{31}^{2} |U_{e1}^{}|^2 \left( 1 - |U_{e 1}^{}|^2 \right)
\right]\;;
\end{eqnarray}
and
\begin{eqnarray}
{\cal S}_{\triangle_{\mu\tau}^{}}^{1} \hspace{-0.17cm} & \simeq &\hspace{-0.17cm}
\frac{\epsilon |U_{e1}^{}|^2}{\Delta_{21}^{} \Delta_{31}^{}}
\left[ \left(m_2^2 m_3^2 - m_1^4\right) \left(1 - |U_{e1}^{}|^2\right)
- m_1^2 \Delta_{32}^{}\left(|U_{e2}^{}|^2 - |U_{e3}^{}|^2\right)
\right]\;,\nonumber\\
{\cal S}_{\triangle_{\mu\tau}^{}}^{2} \hspace{-0.17cm} & \simeq &\hspace{-0.17cm}
-\frac{\epsilon |U_{e2}^{}|^2}{\Delta_{21}^{} \Delta_{32}^{}}
\left[ \left(m_1^2 m_3^2 - m_2^4\right) \left(1 - |U_{e2}^{}|^2\right)
- m_2^2 \Delta_{31}^{}\left(|U_{e1}^{}|^2 - |U_{e3}^{}|^2\right)
\right]\;,\nonumber\\
{\cal S}_{\triangle_{\mu\tau}^{}}^{3} \hspace{-0.17cm} & \simeq &\hspace{-0.17cm}
\frac{\epsilon |U_{e3}^{}|^2}{\Delta_{31}^{} \Delta_{32}^{}}
\left[ \left(m_1^2 m_2^2 - m_3^4\right) \left(1 - |U_{e3}^{}|^2\right)
- m_3^2 \Delta_{21}^{}\left(|U_{e1}^{}|^2 - |U_{e2}^{}|^2\right)
\right]\;,
\end{eqnarray}
where $|U_{\mu i}|^2=|U_{\tau i}|^2$ has been be replaced by 
$(1-|U_{e i}|^2)/2$. We can see that ${\cal S}^{1}_{\triangle_{\mu\tau}^{}}$
and ${\cal S}^{2}_{\triangle_{\mu\tau}^{}}$ are most sensitive
to the neutrino mass ordering. The absolute values of
${\cal S}^{\mu\tau}_{\triangle_{3}^{}}$ and
${\cal S}^{3}_{\triangle_{\mu\tau}^{}}$ should be smaller
because of the smallness of $\Delta_{21}^{}$ and $|U_{e3}^{}|^2$.
The Jarlskog invariant ${\cal J}^{\prime}$ at $\Lambda_{\rm EW}^{}$
running from $\Lambda_{\mu\tau}^{}$ can be written as
\begin{eqnarray}
{\cal J}^{\prime} \hspace{-0.17cm} & \simeq &\hspace{-0.17cm}
{\cal J} - \frac{\epsilon {\cal J}}{\Delta_{21}^{}
	\Delta_{31}^{} \Delta_{32}^{} } \left[ m_1^{2}
\left( m_2^{4} + m_3^{4}\right) \left(|U_{e 2}^{}|^2
-|U_{e 3}^{}|^2 \right) - m_2^{2}
\left( m_1^{4} + m_3^{4}\right) \left(|U_{e 1}^{}|^2
-|U_{e 3}^{}|^2 \right) \right.\nonumber\\&&\hspace{-0.17cm}
\left. + m_3^{2}
\left( m_1^{4} + m_2^{4}\right) \left(|U_{e1 }^{}|^2
-|U_{e2}^{}|^2 \right)\right] \;,
\end{eqnarray}
whose magnitude is proportional to the area of the LUTs 
at $\Lambda_{\rm EW}^{}$. Taking account of
\begin{eqnarray}
\cot \phi_{\alpha i}^{\prime} - \cot \phi_{\beta i}^{\prime}
\hspace{-0.17cm} & = &\hspace{-0.17cm}
\frac{\eta_{\phi}^{}}{\cal J^{\prime}} \left( |U_{\alpha j}^{\prime}
U_{\alpha k}^{\prime *}|^2 - |U_{\beta j}^{\prime}
U_{\beta k}^{\prime *}|^2 \right)\;,
\end{eqnarray}
one obtains $(\cot \phi_{\tau i}^{\prime} 
- \cot \phi_{\mu i}^{\prime})
\simeq -\eta_{\phi}^{} S_{\triangle_i^{}}^{\mu\tau} /{\cal J}$, where
only the first order of $\epsilon$ is kept and
$S_{\triangle_i^{}}^{\mu\tau}$ have been shown in Eq. (44).
Noticing that
bigger $\phi_{\mu i}^{\prime} - \phi_{\tau i}^{\prime} $ lead to
bigger $(\cot \phi_{\tau i}^{\prime} 
- \cot \phi_{\mu i}^{\prime})$ and $S_{\triangle_i^{}}^{\mu\tau}$,
we can also use the more intuitive asymmetries 
$\phi_{\mu i}^{\prime} - \phi_{\tau i}^{\prime} $ to replace 
$S_{\triangle_i^{}}^{\mu\tau}$. The asymmetries of these three 
pairs of inner angles 
satisfy $\sum_i^{}(\phi_{\mu i}^{\prime} 
- \phi_{\tau i}^{\prime}) =0$. 
\subsection{The case of Majorana neutrinos}
When it comes to the Majorana neutrinos, 
the $\mu$-$\tau$ reflection 
symmetry implies the effective Majorana mass term should stay 
unchanged under the flavor and charge-conjugation transformations 
of neutrino fields: $\nu^{}_{e \rm L} 
\leftrightarrow \nu^{\rm c}_{e \rm R}$,
$\nu^{}_{\mu \rm L} \leftrightarrow \nu^{\rm c}_{\tau \rm R}$ and
$\nu^{}_{\tau \rm L} \leftrightarrow \nu^{\rm c}_{\mu \rm R}$.
This results in the limits to the elements of 
neutrino mass matrix $M_{\nu}^{}$:
$(M_{\nu}^{})_{ee}^{} = (M_{\nu}^{})_{ee}^{*}$, 
$(M_{\nu}^{})_{e \mu}^{} = (M_{\nu}^{})_{e \tau}^{*}$,
$(M_{\nu}^{})_{\mu \mu}^{} = (M_{\tau}^{})_{\tau \tau}^{*}$ and
$(M_{\nu}^{})_{\mu \tau}^{} = (M_{\tau}^{})_{\mu \tau}^{*}$
with $(M_{\nu}^{})_{\alpha \beta}^{} \equiv m_i^{} 
U_{\alpha i}^{} U_{\beta i}^{}$ 
being defined in subsection 2.2. Furthermore,
the constraint conditions can be expressed as 
$U_{ei}^{} = \eta_{i}^{} U_{ei}^{*}$ and 
$U_{\mu i}^{} = \eta_{i}^{} U_{\tau i}^{*}$ with $\eta_{i}^{} = 
\pm 1$. Four 
of the eight choices of $(\eta_1^{},\eta_2^{},\eta_3^{})$
are independent
because we can not blackefine the Majorana neutrino fields to change
the sign of arbitrary column of $U$ just like the Dirac case.
Given the $\mu$-$\tau$ reflection symmetry at $\Lambda_{\mu\tau}$,
one gets $|U_{\mu i}^{}|=|U_{\tau i}^{}|$, which results in three isosceles LUTs $\triangle_i^{}$
with $|U_{\mu j}^{} U_{\mu k}^{*}|=|U_{\tau j}^{} U_{\tau k}^{*}|$ 
and a pair of congruent triangles $\left(\triangle_{\mu}^{}\right.$ and 
$\left.\triangle_{\tau}^{}\right)$ with $|U_{\tau i}^{} U_{e i}^{*}|=
|U_{e i}^{} U_{\mu i}^{*}|$  just as
the Dirac case.
So the asymmetries defined in
Eqs. (42) and (43) can be used to denote the deviations
of LUTs of the
Majorana neutrinos at $\Lambda_{\rm EW}^{}$ 
from their special shapes
at $\Lambda_{\mu\tau}^{}$. The  
analytical approximations of 
these asymmetries in this case can be obtained with the
help of $U_{\mu i}^{}=
\eta_{i}^{} U_{\tau i}^{*}$ and Eqs. (31)---(33).
The results are
\begin{eqnarray}
{\cal S}_{\triangle_1^{}}^{\mu\tau} \hspace{-0.17cm} & \simeq 
&\hspace{-0.17cm}
\frac{\epsilon}{\Delta_{21}^{} \Delta_{31}^{} \Delta_{32}^{}}
\left[ \eta_2^{} m_2^{} \Delta_{31}^{} \left(
\eta_2^{} m_2^{}\Delta_{31}^{} -\eta_1^{} m_1^{} \Delta_{32}^{} 
\right)
|U_{e3}^{}|^2 \left(1 - |U_{e 3}^{}|^2\right) 
- \eta_3^{} m_3^{} \Delta_{21}^{} \left(
\eta_1^{} m_1^{} \Delta_{32}^{} \right.\right.\nonumber\\
&&\left.\left.\hspace{-0.17cm} + \eta_3^{} m_3^{} \Delta_{21}^{}\right)
|U_{e2}^{}|^2 \left(1 - |U_{e 2}^{}|^2\right)  
+ \eta_2^{} \eta_3^{} m_2^{} m_3^{}
\Delta_{21}^{} \Delta_{31}^{} \left(|U_{e 2}^{}|^2 -
|U_{e 3}^{}|^2\right) |U_{e 1}^{}|^2
\right]\;,\nonumber\\
{\cal S}_{\triangle_2^{}}^{\mu\tau} \hspace{-0.17cm} & \simeq 
&\hspace{-0.17cm}
\frac{\epsilon}{\Delta_{21}^{} \Delta_{31}^{} \Delta_{32}^{}}
\left[\eta_3^{} m_3^{} \Delta_{21}^{} \left(
\eta_3^{} m_3^{} \Delta_{21}^{} - \eta_2^{} m_2^{} \Delta_{31}^{}\right)
|U_{e1}^{}|^2 \left(1 - |U_{e 1}^{}|^2\right) 
+\eta_1^{} m_1^{} \Delta_{32}^{} \left(
\eta_2^{} m_2^{} \Delta_{31}^{} \right.\right.\nonumber\\
&&\left.\left.\hspace{-0.17cm} - \eta_1^{} m_1^{} \Delta_{32}^{}\right)
|U_{e3}^{}|^2 \left(1 - |U_{e 3}^{}|^2\right) + \eta_1^{} \eta_3^{} m_1^{} m_3^{}
\Delta_{21}^{} \Delta_{32}^{} \left(|U_{e 1}^{}|^2 -
|U_{e 3}^{}|^2\right) |U_{e 2}^{}|^2
\right]\;,\nonumber\\
{\cal S}_{\triangle_3^{}}^{\mu\tau} \hspace{-0.17cm} & \simeq 
&\hspace{-0.17cm}
\frac{\epsilon}{\Delta_{21}^{} \Delta_{31}^{} \Delta_{32}^{}}
\left[ \eta_1^{} m_1^{} \Delta_{32}^{} \left(
\eta_1^{} m_1^{} \Delta_{32}^{} + \eta_3^{} m_3^{} \Delta_{21}^{}\right)
|U_{e2}^{}|^2 \left( 1 - |U_{e 2}^{}|^2 \right)
+\eta_2^{} m_2^{} \Delta_{31}^{} \left(
\eta_3^{} m_3^{} \Delta_{21}^{} \right.\right.\nonumber\\
&&\left.\left.\hspace{-0.17cm} - \eta_2^{} m_2^{} \Delta_{31}^{}\right)
|U_{e1}^{}|^2 \left(1- |U_{e 1}^{}|^2\right) +\eta_1^{} \eta_2^{} m_1^{} m_2^{}
\Delta_{31}^{} \Delta_{32}^{} \left(|U_{e 1}^{}|^2 -
|U_{e 2}^{}|^2\right) |U_{e 3}^{}|^2
\right]\;,
\end{eqnarray}
demonstrating the deviations of
$\triangle_i^{\prime}$ at $\Lambda_{\rm EW}^{}$ from their 
isosceles shapes
at $\Lambda_{\mu \tau}^{}$; and
\begin{eqnarray}
{\cal S}_{\triangle_{\mu\tau}^{}}^{1} \hspace{-0.17cm} & \simeq &\hspace{-0.17cm}
\frac{\epsilon |U_{e1}^{}|^2}{\Delta_{21}^{} \Delta_{31}^{}}
\left[ \left(m_2^2 m_3^2 - m_1^4\right) \left(1 - |U_{e1}^{}|^2\right)
- m_1^2 \Delta_{32}^{}\left(|U_{e2}^{}|^2 - |U_{e3}^{}|^2\right)
\right. \nonumber\\ && \hspace{-0.17cm} \left.-2 \eta_1^{} m_1^{}
\left(\eta_2^{} m_2^{} \Delta_{31}^{}  |U_{e3}^{}|^2 + \eta_3^{} 
m_3^{} \Delta_{21}^{}  |U_{e2}^{}|^2\right)
\right]\;,\nonumber\\
{\cal S}_{\triangle_{\mu\tau}^{}}^{2} \hspace{-0.17cm} & \simeq &\hspace{-0.17cm}
-\frac{\epsilon |U_{e2}^{}|^2}{\Delta_{21}^{} \Delta_{32}^{}}
\left[ \left(m_1^2 m_3^2 - m_2^4\right) \left(1 - |U_{e2}^{}|^2\right)
- m_2^2 \Delta_{31}^{}\left(|U_{e1}^{}|^2 - |U_{e3}^{}|^2\right)
\right. \nonumber\\ && \hspace{-0.17cm} \left.-2 \eta_2^{} m_2^{}
\left(\eta_1^{} m_1^{} \Delta_{32}^{}  |U_{e3}^{}|^2 - \eta_3^{} 
m_3^{} \Delta_{21}^{}  |U_{e1}^{}|^2\right)
\right]\;,\nonumber\\
{\cal S}_{\triangle_{\mu\tau}^{}}^{3} \hspace{-0.17cm} & \simeq &\hspace{-0.17cm}
\frac{\epsilon |U_{e3}^{}|^2}{\Delta_{31}^{} \Delta_{32}^{}}
\left[ \left(m_1^2 m_2^2 - m_3^4\right) \left(1 - |U_{e3}^{}|^2\right)
- m_3^2 \Delta_{21}^{}\left(|U_{e1}^{}|^2 - |U_{e2}^{}|^2\right)
\right. \nonumber\\ && \hspace{-0.17cm} \left.+2 \eta_3^{} m_3^{}
\left(\eta_1^{} m_1^{} \Delta_{32}^{}  |U_{e2}^{}|^2 + \eta_2^{} 
m_2^{} \Delta_{31}^{} |U_{e1}^{}|^2\right)
\right]\;,
\end{eqnarray}
showing the deviations of  
$\triangle_{\mu}^{\prime}$ and $\triangle_{\tau}^{\prime}$
at $\Lambda_{\rm EW}^{}$ from their congruent shapes
at $\Lambda_{\mu \tau}^{}$.
From Eqs. (48) and (49),
we find that ${\cal {S}}^{1}_{\triangle_{\mu\tau}^{}}$ and 
${\cal {S}}^{2}_{\triangle_{\mu\tau}^{}}$ are most sensitive 
to the neutrino mass ordering; ${\cal 
{S}}^{3}_{\triangle_{\mu\tau}^{}}$ and ${\cal 
{S}}^{\mu\tau}_{\triangle_3^{}}$ are smaller
due to the suppression of $\Delta_{21}^{}$ and $|U_{e3}^{}|^2$.
This conclusion is the same as the Dirac case.
The connection of the Jarlskog invariants of Majorana neutrinos between 
$\Lambda_{\rm EW}^{}$ and $\Lambda_{\mu \tau}^{}$ can be written as
\begin{eqnarray}
{\cal J}^{\prime} \hspace{-0.17cm} & \simeq &\hspace{-0.17cm}
{\cal J} - \frac{\epsilon {\cal J}}{\Delta_{21}^{}
	\Delta_{31}^{} \Delta_{32}^{} } \left\{ \left[m_1^{2}
\left( m_2^{4} + m_3^{4}\right)-\eta_2^{}
\eta_3^{} m_2^{} m_3^{} \Delta_{21}^{} \Delta_{31}^{}\right] 
\left(|U_{e 2}^{}|^2-|U_{e 3}^{}|^2 \right) 
\right.\nonumber\\&&\hspace{-0.17cm} \left.
- \left[m_2^{2} \left( m_1^{4} + m_3^{4}\right) + 
\eta_1^{}\eta_3^{} m_1^{} m_3^{} \Delta_{21}^{}\Delta_{32}^{}
\right]\left(|U_{e 1}^{}|^2-|U_{e 3}^{}|^2 \right)
\right.\nonumber\\&&\hspace{-0.17cm} \left. 
+ \left[m_3^{2} \left( m_1^{4} + m_2^{4}\right) 
-\eta_1^{} \eta_2^{} m_1^{} m_2^{} \Delta_{31}^{}
\Delta_{32}^{}\right]
\left(|U_{e1 }^{}|^2 - |U_{e2}^{}|^2 \right) \right\}\;.
\end{eqnarray}
From Eqs. (47), (48) and (50), we can get $(\cot \phi_{\tau 
i}^{\prime} - \cot \phi_{\mu i}^{\prime})\simeq -\eta_{\phi}^{} 
S_{\triangle_i^{}}^{\mu\tau} /{\cal J}$. The magnitude
of $(\cot \phi_{\tau i}^{\prime} - \cot \phi_{\mu i}^{\prime})$
always keeps consistent with that of $\phi_{\mu i}^{\prime} 
- \phi_{\tau i}^{\prime}$ or $S_{\triangle_i^{}}^{\mu\tau}$.

It is clear to see that the analytical approximations of 
${\cal {S}}^{\mu\tau}_{\triangle_i^{}}$, ${\cal 
S}^i_{\triangle_{\mu\tau}^{}}$ and ${\cal J}^{\prime}$ 
for the Majorana neutrinos include more
odd terms of $\eta_i^{}$ (i.e., $\eta_i^{}\eta_j^{}$ 
for $i \neq j$)
compablack with their counterparts for the Dirac neutrinos.
These terms can be directly connected with the Majorana phases and
have complicated influence on the LUT reformations at
$\Lambda_{\rm EW}^{}$.
 
\section{Numerical analysis}
Before we start the numerical analysis, let us first parametrize
$U$ as 
\begin{eqnarray}
U= \left(\begin{matrix}
c^{}_{12} c^{}_{13} & s^{}_{12} c^{}_{13} &
s^{}_{13} e^{-{\rm i} \delta} \cr -s^{}_{12} c^{}_{23} - c^{}_{12}
s^{}_{13} s^{}_{23} e^{{\rm i} \delta} & c^{}_{12} c^{}_{23} -
s^{}_{12} s^{}_{13} s^{}_{23} e^{{\rm i} \delta} & c^{}_{13}
s^{}_{23} \cr -s^{}_{12} s^{}_{23} + c^{}_{12} s^{}_{13} c^{}_{23}
e^{{\rm i} \delta} & c^{}_{12} s^{}_{23} + s^{}_{12} s^{}_{13}
c^{}_{23} e^{{\rm i} \delta} &  - c^{}_{13} c^{}_{23} \cr
\end{matrix} \right) \;
\end{eqnarray}
for the Dirac neutrinos with $c_{ij}^{} \equiv \cos \theta_{ij}^{}$ and $s_{ij}^{} \equiv
\sin \theta_{ij}^{}$. For the Majorana neutrinos,
one has to add the Majorana phase matrix
$P_{\nu}^{} \equiv {\rm Diag} \left\{ e^{i\rho}, 
e^{i\sigma}, 1 \right\}$ on the right side of Eq. (51).
$U^{\prime}$ at $\Lambda_{\rm EW}^{}$ has the same form as
$U$ with the corresponding set of flavor
mixing angles and CP phases $(\theta_{12}^{\prime},\theta_{13}^{\prime},
\theta_{23}^{\prime},\delta^{\prime},\rho^{\prime},\sigma^{\prime})$.
According to the specific parametrization of $U$ in Eq. (51),
we interpret the constraints of the 
$\mu$-$\tau$ reflection symmetry 
as two conditions for the Dirac neutrinos: 
$\theta_{23}^{}=\pi/4$ and
$\delta=\pm \pi/2$, and four 
conditions for the Majorana neutrinos: $\theta_{23}^{}=\pi/4$, $\delta=\pm \pi/2$,
$\rho=0$ or $\pi/2$ and $\sigma=0$ or $\pi/2$.  
The correspondences between the eight choices of $(\delta,\rho,\sigma)$ 
and the four independent cases of $(\eta_1^{},\eta_2^{},\eta_3^{})$ 
have been listed in Table 1.
Given the fact that the global-fit analysis of current 
neutrino oscillation data has
implied a preference of $\delta$ around $-\pi/2$ \cite{Lisi,Valle},
we only focus on the case $\delta=-\pi/2$ at $\Lambda_{\mu\tau}^{}$ 
for both Dirac and Majorana neutrinos. The framework
of the MSSM is typically chosen because the RGE-induced $\mu$-$\tau$ reflection
symmetry breaking is always very small in the SM \cite{Ohlsson:2013xva}. 
\begin{table}[H]
	\caption{The correspondences between $(\delta,\rho,\sigma)$
    and $(\eta_1^{},\eta_2^{}, \eta_3^{})$ in the $\mu$-$\tau$
    reflection symmetry limit for the Majorana neutrinos.}
	\vspace{0.3cm}
	\centering
	\renewcommand\arraystretch{1.1}
\begin{tabular}{c|c} \hline\hline
$\left(\delta, \rho, \sigma \right)$ & $(\eta_1^{}, \eta_2^{}, \eta_3^{})$ \\
\hline 
$\left(\pm \frac{\pi}{2},0,0\right)$ & $(1, 1,-1)$ \\ \hline
$\left(\pm \frac{\pi}{2},\frac{\pi}{2},0\right)$ & $(-1, 1,-1)$ \\ \hline
$\left(\pm \frac{\pi}{2}, 0, \frac{\pi}{2}\right)$ & $(1,-1,-1)$ \\ \hline
$\left(\pm \frac{\pi}{2},\frac{\pi}{2},\frac{\pi}{2}\right)$ 
& $(-1, -1,-1)$ \\ \hline\hline
\end{tabular}
\end{table}
To show the deviations of the six LUTs at $\Lambda_{\rm EW}^{}$ 
from their special shapes at $\Lambda_{\mu\tau}^{}$,
which can be described by the asymmetries defined in section 3,
the numerical analysis similar to that in Ref. \cite{Huang}
has been done. Both the NMO $(m_1^{(\prime)} < m_2^{(\prime)} 
< m_3^{(\prime)})$ and IMO $(m_3^{(\prime)} < m_1^{(\prime)} 
< m_2^{(\prime)})$ cases of the Dirac or Majorana
neutrinos will be taken into account.
Note that there are four choices of
the two Majorana phases at $\Lambda_{\mu\tau}^{}$, which
need to be consideblack separately, too. 
In each case, 
we first run the relevant RGEs from 
$\Lambda_{\mu\tau}^{} \sim 10^{14} {\rm GeV}$ down to
$\Lambda_{\rm EW}^{}\sim 10^{2} {\rm GeV}$ in the framework
of MSSM. 
{\color {black} {
Here we roughly take the MSSM breaking scale 
$\Lambda_{\rm MSSM}^{}$ around $\Lambda_{\rm EW}^{}$.
\footnote{The $\Lambda_{\rm MSSM}^{}$, where all superpartners are
integrated out at once, is just
the matching scale of SM and MSSM. It is usually assumed to be
around the MSSM particle mass scales, i.e., from 1 TeV to 10 TeV. 
Because the range from $\Lambda_{\rm MSSM}^{}$ to $\Lambda_{\rm EW}^{}$
is much smaller than the one from $\Lambda_{ \mu\tau}^{}$ to $\Lambda_{\rm EW}^{}$ and the RGE running effect on neutrino mass parameters from $\Lambda_{\rm MSSM}^{}$ to $\Lambda_{\rm EW}^{}$ is very small, we can
roughly take $\Lambda_{\rm MSSM}^{} \simeq \Lambda_{\rm EW}^{}$.
}.
In the case of the Majorana neutrinos, we 
assume 
that all the heavy singlet neutrinos have a 
mass spectrum at $\Lambda_{\mu\tau}^{}$ and are all integrated out at $\Lambda_{\mu\tau}^{}$ 
\footnote {If the heavy neutrino masses are below
$\Lambda_{\mu\tau}^{}$, we need to integrate out them successively and take into account different effective theories corresponding to different ranges of the renormalization energy scale. Thus the final results of neutrino
mass parameters at $\Lambda_{\rm EW}^{}$ running from $\Lambda_{\mu\tau}^{}$ may be very different from our scenario under consideration \cite{Antusch:2005gp,Nath:2018hjx,Ohlsson:2013xva,Antusch:2002rr,Mei:2004rn}.}.
}}
The initial values  at $\Lambda_{\mu\tau}^{}$ include the corresponding $\mu$-$\tau$ reflection symmetry 
constraint conditions of flavor mixing angles and CP phases.
Furthermore, the smallest neutrino mass ($m_1^{\prime}$ for
the NMO case and $m_3^{\prime}$ for the IMO case) at $\Lambda_{\rm EW}^{}$
and the MSSM parameter $\tan\beta$ vary in the reasonable 
ranges $[0,0.1]$ eV and $[10,50]$, respectively. 
For each given values of $m_1^{\prime}$
or $m_3^{\prime}$  and $\tan\beta$, the other parameters $(\sin^2\theta_{12}^{},
\sin^2\theta_{13}^{}, \Delta_{\rm sol}^{}, \Delta_{\rm atm}^{})$ at
$\Lambda_{\mu\tau}^{}$ are scanned over wide enough ranges by means of the 
MultiNest program \cite{Feroz}, where $\Delta_{\rm sol}^{}
=m_2^2 - m_1^2$, $\Delta_{\rm atm}^{}= m_3^2 - (m_1^2 + m_2^2)/2$,
and their counterparts at $\Lambda_{\rm EW}^{}$
$\Delta_{\rm sol}^{\prime}=m_2^{\prime 2} - m_1^{\prime 2}$ and 
$\Delta_{\rm atm}^{\prime}= m_3^{\prime 2} - (m_1^{\prime 2} 
+ m_2^{\prime 2})/2$ 
have been defined to keep consistent with the notations in Ref. \cite{Lisi}. 
From each scan, we can
get a set of parameters at $\Lambda_{\rm EW}^{}$
which will be confronted with the latest global-fit results of
current neutrino oscillation data by
\begin{eqnarray}
\chi^2 \equiv \sum_{i=1}^{6} \frac{\left(\xi_i - 
\overline{\xi}_i\right)^2}{\sigma_i^2} \; ,
\end{eqnarray}
where $\xi_i^{} \in \{\sin^2\theta_{12}^{\prime}, \sin^2\theta_{13}^{\prime}, 
\sin^2\theta_{23}^{\prime},\delta^{\prime},\Delta_{\rm sol}^{\prime}, 
\Delta_{\rm atm}^{\prime}\}$ stand for the oscillation 
parameters yielded 
from the scan; $\overline \xi_i^{}$ and $\sigma_i^{}$ denote 
the best-fit values and averaged $1\sigma$ errors
of $\xi_i^{}$ from the global-fit analysis
in Ref. \cite{Lisi}, respectively.
The best-fit values and $3\sigma$ ranges of 
${\cal S}^{i}_{\triangle_{\mu\tau}^{}}$, $\phi_{\mu i}^{\prime}
- \phi_{\tau i}^{\prime}$ and ${\cal J}^{\prime}$ 
are listed in Tables 2---6,
corresponding to the minimal values $\chi_{\rm min}^{2}$ of $\chi^2$
and $\chi^2 \le 9$ for one degree of freedom, respectively.
Considering that the two asymmetries 
${\cal S}^{\mu\tau}_{\triangle_{i}^{}}$ and 
$\phi_{\mu i}^{\prime} - 
\phi_{\tau i}^{\prime}$ imply consistent deviations of the LUTs, 
we only demonstrate the numerical 
results of the latter.
Some discussions about the numerical results are as follows:
\begin{itemize}
\item 
Complementary to the
analytical approximations in section 3, the numerical results 
generally reveal how the six LUTs can be reformed
at $\Lambda_{\rm EW}^{}$ by assuming the 
$\mu$-$\tau$ reflection symmetry
at $\Lambda_{\mu\tau}^{}$. The reformations depend a lot on
the lightest neutrino mass, the neutrino mass ordering, 
the Majorana phases and $\tan\beta$. From Tables 2---6,
we find that the parameters running from 
$\Lambda_{\mu\tau}^{}$ 
and their corresponding best-fit values
from the global analysis in Ref. \cite{Lisi} can not fit very well 
in the IMO case, leading to big values of $\chi^2_{\rm min}$.
This is mainly because the running direction of $\theta_{23}^{}$
from $\Lambda_{\mu\tau}^{}$ to $\Lambda_{\rm EW}^{}$ is opposite
to its best-fit value in this case
\cite{Huang,Xing:2017mkx}. The
lightest neutrino mass $m_3^{\prime}$ and $\tan\beta$ are
limited to smaller ranges by $\chi^2_{} \le 9$ 
in Tables 3---5. 
\item 
The deviations of the six LUTs are small for the case
$(\rho,\sigma)= (\pi/2, \pi/2)$ in Table 6 but their values can
be very big in some other cases. For example, the two asymmetries $\phi^{\prime}_{\mu 1} - \phi^{\prime}_{\tau 1}$ 
and 
$\phi^{\prime}_{\mu 2} - \phi^{\prime}_{\tau 2}$ may reach about
$180^{\circ}$ in magnitude because of the smallness of 
the corresponding $\cal J^{\prime}$. We also notice that ${\cal J}^{\prime}$
running from $\Lambda_{\mu\tau}^{}$ can not be zero due
to the nonzero value of $\cal J^{}$ constrained by the $\mu$-$\tau$
reflection symmetry conditions. It is easy to understand
this point from Eqs. (46) and (50).
\item
The smallest $\chi^2_{\rm min}$ for the 
Dirac and Majorana neutrinos
come from the best-fit results 
of the NMO case in Table 2 and Table 5, respectively. 
The corresponding LUTs together with their counterparts
at $\Lambda_{\mu\tau}^{}$ have been 
specifically shown in Fig. 1 and Fig. 2.
The blue triangles with $\chi_{\rm min}^2 \simeq 0.01$ 
stand for the LUTs at $\Lambda_{\rm EW}^{}$ 
and almost overlap the
LUTs implied by the best-fit values of the global analysis
in Ref. \cite{Lisi}, while the black ones denote the 
corresponding LUTs at $\Lambda_{\mu\tau}^{}$. 
When comparing the two figures, we find that 
the blue LUTs at $\Lambda_{\rm EW}^{}$ differ with each other
only in the orientations of $\triangle_{i}^{}$ caused by the Majorana phases,
while the black ones are very different.
\end{itemize}
\begin{table}[H]
\caption{The numerical analysis of deviations of
	the six LUTs  at $\Lambda_{\rm EW}^{}$ from
	their $\mu$-$\tau$ reflection symmetry limits at $\Lambda_{\mu\tau}^{}$ for the
	Dirac neutrinos in the framework of the MSSM, by inputting  
	$(\theta_{23}^{},\delta) =(\pi/4,-\pi/2)$ at $\Lambda_{\mu\tau}^{}$
	and allowing the smallest neutrino mass 
	($m_1^{\prime}$ for the NMO case and
    $m_3^{\prime}$ for the IMO case) and the MSSM parameter
	$\tan \beta$ to vary in the ranges $[0,0.1]$ eV
	and $[10,50]$, respectively. }
	\vspace{0.2cm}
	\centering
	\renewcommand\arraystretch{1.1}
\begin{tabular}{lllllllll} \hline \hline
	&& \multicolumn{3}{l}{Normal mass ordering (NMO)} &&
	\multicolumn{3}{l}{Inverted mass ordering (IMO)} \\ \hline
	&& \hspace{-0.2cm}$\begin{array}{l} {\hspace{-0.12cm}
	\text{ best-fit}} \\ \chi_{\rm min}^2 \simeq 0.01 \end{array}$ 
    && \hspace{-0.2cm}$\begin{array}{l} 3\sigma {\text{ range}} 
    \\\chi_{}^2 \le 9\end{array}$
    && \hspace{-0.2cm}$\begin{array}{l} {\hspace{-0.12cm}
    \text{ best-fit}} \\ \chi_{\rm min}^2 \simeq 7.94 \end{array}$ 
    && \hspace{-0.2cm}$\begin{array}{l} 3\sigma {\text{ range}} 
    \\\chi_{}^2 \le 9\end{array}$ \\ \hline
	$S^{1}_{\triangle_{\mu\tau}^{}}/10^{-2}$ && $6.60$ && 
	$(0.03, 13.73)$ && $0.11$ && $(0.10,0.64)$ \\
    $S^{2}_{\triangle_{\mu\tau}^{}}/10^{-2}$ && $0.25$ && 
    $(-0.15, 1.34)$ && $-0.08$ && $(-0.52,-0.08)$ \\
	$S^{3}_{\triangle_{\mu\tau}^{}}/10^{-3}$ && $-2.26$ && 
	$(-5.16, -0.03)$ && $0.03$ && $(0.03,0.18)$ \\
	\hline 
	$\phi_{\mu 1}^{\prime}-\phi_{\tau 1}^{\prime}$ && $59.45^{\circ}$
	&&$(0.03^{\circ},178.71^{\circ})$ && $1.60^{\circ}$ && 
	$(1.55^{\circ}, 9.79^{\circ})$  \\
	$\phi_{\mu 2}^{\prime}-\phi_{\tau 2}^{\prime}$ && $-54.09^{\circ}$
	&&$(-178.53^{\circ},0.60^{\circ})$ && $-1.58^{\circ}$ 
	&& $(-9.70^{\circ},-1.54^{\circ})$  \\
	$\phi_{\mu 3}^{\prime}-\phi_{\tau 3}^{\prime}$ &&$-5.36^{\circ}$ 
	&& $(-8.20^{\circ},-0.05^{\circ})$ && $-0.02^{\circ}$ 
	&& $(-0.10^{\circ},0.07^{\circ})$ 
	\\  \hline 
	${\cal{J^{\prime}}}/10^{-2}$&& $-2.85$  && $(-3.48,-0.04)$ && $-3.32$ && 
	$(-3.39,-3.25)$ \\ \hline	
	$m_1^{\prime}$ or $m_3^{\prime}$/eV && $0.085$  && $(0,0.1)$ 
	&& $0.001$ && 
	$(0,0.078)$ \\ \hline
	$\tan\beta$&& $32$  && $(10,50)$ && $10$ && 
	$(10,24)$ \\
	\hline\hline
\end{tabular}
\vspace{0.2cm}
\end{table}
\begin{table}[H]
	\caption{The numerical analysis of deviations of
		the six LUTs at $\Lambda_{\rm EW}^{}$ from
		their $\mu$-$\tau$ reflection symmetry limits at $\Lambda_{\mu\tau}^{}$ for
		the Majorana neutrinos in the framework of the MSSM, by inputting  
		$(\theta_{23}^{},\delta,\rho,\sigma) =(\pi/4,-\pi/2,0,0)$ 
		at $\Lambda_{\mu\tau}^{}$
		and allowing the smallest neutrino mass at $\Lambda_{\rm EW}^{}$
		($m_1^{\prime}$ for the NMO case and
	$m_3^{\prime}$ for the IMO case) and the MSSM parameter
		$\tan \beta$ to vary in the ranges $[0,0.1]$ eV
		and $[10,50]$, respectively.}
	\vspace{0.2cm}
	\centering
	\renewcommand\arraystretch{1.1}
	\begin{tabular}{lllllllll} \hline\hline
	&& \multicolumn{3}{l}{Normal mass ordering (NMO)} &&
		\multicolumn{3}{l}{Inverted mass ordering (IMO)} \\ \hline
	&& \hspace{-0.2cm}$\begin{array}{l} {\hspace{-0.12cm}
	\text{ best-fit}} \\ \chi_{\rm min}^2 \simeq 0.77 \end{array}$ 
&& \hspace{-0.2cm}$\begin{array}{l} 3\sigma {\text{ range}} 
\\\chi_{}^2 \le 9\end{array}$
&& \hspace{-0.2cm}$\begin{array}{l} {\hspace{-0.12cm}
	\text{ best-fit}} \\ \chi_{\rm min}^2 \simeq 7.94 \end{array}$ 
&& \hspace{-0.2cm}$\begin{array}{l} 3\sigma {\text{ range}} 
\\\chi_{}^2 \le 9\end{array}$ \\ \hline
		$S^{1}_{\triangle_{\mu\tau}^{}}/10^{-2}$ && $2.03$ && 
		$(0.03, 6.75)$ && $-0.03$ && $(-0.18,-0.03)$ \\
		$S^{2}_{\triangle_{\mu\tau}^{}}/10^{-2}$ && $2.10$ && 
		$(0.04, 7.02)$ && $-0.03$ && $(-0.18,-0.03)$ \\
		$S^{3}_{\triangle_{\mu\tau}^{}}/10^{-3}$ && $-2.15$ && 
		$(-7.25, -0.03)$ && $0.03$ && $(0.03,0.19)$ \\
		\hline 
		$\phi_{\mu 1}^{\prime}-\phi_{\tau 1}^{\prime}$ && $1.23^{\circ}$
		&&$(-1.95^{\circ},4.30^{\circ})$ && $-0.02^{\circ}$ && 
		$(-0.19^{\circ},-0.02^{\circ})$  \\
		$\phi_{\mu 2}^{\prime}-\phi_{\tau 2}^{\prime}$ && $2.39^{\circ}$
		&&$(0.08^{\circ},8.56^{\circ})$ && $-0.02^{\circ}$ 
		&& $(-0.15^{\circ},-0.02^{\circ})$  \\
		$\phi_{\mu 3}^{\prime}-\phi_{\tau 3}^{\prime}$ &&$-3.62^{\circ}$ 
		&& $(-12.69^{\circ},-0.06^{\circ})$ && $0.04^{\circ}$ 
		&& $(0.05^{\circ},0.32^{\circ})$ 
		\\  \hline 
		${\cal{J}^{\prime}}/10^{-2}$&& $-3.27$  && $(-3.48,-3.03)$ && $-3.32$ && 
		$(-3.39,-3.25)$
		\\ \hline
		$m_1^{\prime}$ or $m_3^{\prime}$/eV && $0.081$  && $(0,0.1)$ 
		&& $3.5\times 10^{-5}$ && 	$(0,0.053)$ \\ \hline
		$\tan\beta$&& $24$  && $(10,50)$ && $10$ && 
		$(10,24)$ \\
		\hline\hline
	\end{tabular}
	\vspace{0.2cm}
\end{table}
\begin{table}[H]
	\caption{The numerical analysis of deviations of
		the six LUTs at $\Lambda_{\rm EW}^{}$ from
	their $\mu$-$\tau$ reflection symmetry limits at $\Lambda_{\mu\tau}^{}$ for
	the Majorana neutrinos in the framework of the MSSM, by inputting  
	$(\theta_{23}^{},\delta,\rho,\sigma) =(\pi/4,-\pi/2,0,\pi/2)$ 
	at $\Lambda_{\mu\tau}^{}$
	and allowing the smallest neutrino mass at $\Lambda_{\rm EW}^{}$
	($m_1^{\prime}$ for the NMO case and
    $m_3^{\prime}$ for the IMO case) and the MSSM parameter
	$\tan \beta$ to vary in the ranges $[0,0.1]$ eV
	and $[10,50]$, respectively.}
	\vspace{0.2cm}
	\centering
	\renewcommand\arraystretch{1.1}
	\begin{tabular}{lllllllll} \hline\hline
		&& \multicolumn{3}{l}{Normal mass ordering (NMO)} &&
		\multicolumn{3}{l}{Inverted neutrino mass ordering} \\ \hline
&& \hspace{-0.2cm}$\begin{array}{l} {\hspace{-0.12cm}
	\text{ best-fit}} \\ \chi_{\rm min}^2 \simeq0.27 \end{array}$ 
&& \hspace{-0.2cm}$\begin{array}{l} 3\sigma {\text{ range}} 
\\\chi_{}^2 \le 9\end{array}$
&& \hspace{-0.2cm}$\begin{array}{l} {\hspace{-0.12cm}
	\text{ best-fit}} \\ \chi_{\rm min}^2 \simeq7.94 \end{array}$ 
&& \hspace{-0.2cm}$\begin{array}{l} 3\sigma {\text{ range}} 
\\\chi_{}^2 \le 9\end{array}$ \\ \hline
		$S^{1}_{\triangle_{\mu\tau}^{}}/10^{-2}$ && $6.48$ && 
		$(0.03, 10.37)$ && $0.23$ && $(0.22,1.33)$ \\
		$S^{2}_{\triangle_{\mu\tau}^{}}/10^{-2}$ && $-1.19$ && 
		$(-4.46, 0.73)$ && $-0.14$ && $(-0.78,-0.13)$ \\
		$S^{3}_{\triangle_{\mu\tau}^{}}/10^{-3}$ && $-1.18$ && 
		$(-1.54, -0.01)$ && $0.03$ && $(0.03,0.16)$ \\
		\hline 
		$\phi_{\mu 1}^{\prime}-\phi_{\tau 1}^{\prime}$ && $71.91^{\circ}$
		&&$(0.11^{\circ},179.99^{\circ})$ && $3.16^{\circ}$ && 
		$(3.05^{\circ},18.01^{\circ})$  \\
		$\phi_{\mu 2}^{\prime}-\phi_{\tau 2}^{\prime}$ && $-68.10^{\circ}$
		&&$(-179.99^{\circ},-0.07^{\circ})$ && $-3.08^{\circ}$ 
		&& $(-17.56^{\circ},-2.98^{\circ})$  \\
		$\phi_{\mu 3}^{\prime}-\phi_{\tau 3}^{\prime}$ &&$-3.81^{\circ}$ 
		&& $(-4.52^{\circ},-0.00001^{\circ})$ && $-0.08^{\circ}$ 
		&& $(-0.45^{\circ},-0.06^{\circ})$ 
		\\  \hline 
		${\cal{J^{\prime}}}/10^{-2}$&& $-2.66$  && $(-3.48,-8.65\times10^{-6})$ && $-3.32$ && 
		$(-3.38,-3.25)$
		\\
		\hline
		$m_1^{\prime}$ or $m_3^{\prime}$/eV && $0.030$  && $(0,0.1)$ 
		&& $9.7\times 10^{-3}$ && 	$(0,0.097)$ \\ \hline
		$\tan\beta$&& $50$  && $(10,50)$ && $10$ && 
		$(10,22)$ \\
		\hline\hline
	\end{tabular}
	\vspace{0.2cm}
\end{table}
\begin{table}[H]
	\caption{The numerical analysis of deviations of
		the six LUTs at $\Lambda_{\rm EW}^{}$ from
	their $\mu$-$\tau$ reflection symmetry limits at $\Lambda_{\mu\tau}^{}$ for
	the Majorana neutrinos in the framework of the MSSM, by inputting  
	$(\theta_{23}^{},\delta,\rho,\sigma) =(\pi/4,-\pi/2,\pi/2,0)$ 
	at $\Lambda_{\mu\tau}^{}$
	and allowing the smallest neutrino mass at $\Lambda_{\rm EW}^{}$
	($m_1^{\prime}$ for the NMO case and
	$m_3^{\prime}$ for the IMO case) and the MSSM parameter
	$\tan \beta$ to vary in the ranges $[0,0.1]$ eV
	and $[10,50]$, respectively.}
	\vspace{0.2cm}
	\centering
	\renewcommand\arraystretch{1.1}
	\begin{tabular}{lllllllll} \hline\hline
		&& \multicolumn{3}{l}{Normal mass ordering (NMO)} &&
		\multicolumn{3}{l}{Inverted mass ordering (IMO)} \\ \hline
&& \hspace{-0.2cm}$\begin{array}{l} {\hspace{-0.12cm}
	\text{ best-fit}} \\ \chi_{\rm min}^2 \simeq 0.01 \end{array}$ 
&& \hspace{-0.2cm}$\begin{array}{l} 3\sigma {\text{ range}} 
\\\chi_{}^2 \le 9\end{array}$
&& \hspace{-0.2cm}$\begin{array}{l} {\hspace{-0.12cm}
	\text{ best-fit}} \\ \chi_{\rm min}^2 \simeq 7.96 \end{array}$ 
&& \hspace{-0.2cm}$\begin{array}{l} 3\sigma {\text{ range}} 
\\\chi_{}^2 \le 9\end{array}$ \\ \hline
		$S^{1}_{\triangle_{\mu\tau}^{}}/10^{-2}$ && $6.59$ && 
		$(0.03, 14.47)$ && $0.24$ && $(0.23,1.24)$ \\
		$S^{2}_{\triangle_{\mu\tau}^{}}/10^{-2}$ && $0.24$ && 
		$(-0.19, 2.40)$ && $-0.14$ && $(-0.76,-0.14)$ \\
		$S^{3}_{\triangle_{\mu\tau}^{}}/10^{-3}$ && $-2.25$ && 
		$(-6.01, -0.03)$ && $0.03$ && $(0.03,0.17)$ \\
		\hline 
		$\phi_{\mu 1}^{\prime}-\phi_{\tau 1}^{\prime}$ && $59.41^{\circ}$
		&&$(-2.70^{\circ},179.81^{\circ})$ && $3.22^{\circ}$ && 
		$(3.13^{\circ},16.94^{\circ})$  \\
		$\phi_{\mu 2}^{\prime}-\phi_{\tau 2}^{\prime}$ && $-54.07^{\circ}$
		&&$(-179.79^{\circ},4.80^{\circ})$ && $-3.14^{\circ}$ 
		&& $(-16.56^{\circ},-3.05^{\circ})$  \\
		$\phi_{\mu 3}^{\prime}-\phi_{\tau 3}^{\prime}$ &&$-5.34^{\circ}$ 
		&& $(-9.64^{\circ},-0.02^{\circ})$ && $-0.08^{\circ}$ 
		&& $(-0.41^{\circ},-0.06^{\circ})$ 
		\\  \hline 
		${\cal{J^{\prime}}}/10^{-2}$&& $-2.85$  && $(-3.48,-0.006)$ && $-3.32$ && 
		$(-3.38,-3.25)$
		\\
		\hline
		$m_1^{\prime}$ or $m_3^{\prime}$/eV && $0.097$  && $(0,0.1)$ 
		&& $1.3\times10^{-4}$ && 	$(0,0.062)$ \\ \hline
		$\tan\beta$&& $24$  && $(10,50)$ && $10$ && 
		$(10,22)$ \\
		\hline\hline
	\end{tabular}
	\vspace{0.2cm}
\end{table}
\begin{table}[H]
	\caption{The numerical analysis of deviations of
		the six LUTs at $\Lambda_{\rm EW}^{}$ from
	their $\mu$-$\tau$ reflection symmetry limits at $\Lambda_{\mu\tau}^{}$ for
	the Majorana neutrinos in the framework of the MSSM, by inputting  
	$(\theta_{23}^{},\delta,\rho,\sigma) =(\pi/4,-\pi/2,\pi/2,\pi/2)$ 
	at $\Lambda_{\mu\tau}^{}$
	and allowing the smallest neutrino mass at $\Lambda_{\rm EW}^{}$	
	($m_1^{\prime}$ for the NMO case and
	$m_3^{\prime}$ for the IMO case) and the MSSM parameter
	$\tan \beta$ to vary in the ranges $[0,0.1]$ eV
	and $[10,50]$, respectively.}
	\vspace{0.2cm}
	\centering
	\renewcommand\arraystretch{1.1}
	\begin{tabular}{lllllllll} \hline\hline
		&& \multicolumn{3}{l}{Normal mass ordering (NMO)} &&
		\multicolumn{3}{l}{Inverted neutrino mass ordering} \\ \hline
&& \hspace{-0.2cm}$\begin{array}{l} {\hspace{-0.12cm}
	\text{ best-fit}} \\ \chi_{\rm min}^2 \simeq 1.17 \end{array}$ 
&& \hspace{-0.2cm}$\begin{array}{l} 3\sigma {\text{ range}} 
\\\chi_{}^2 \le 9\end{array}$
&& \hspace{-0.2cm}$\begin{array}{l} {\hspace{-0.12cm}
	\text{ best-fit}} \\ \chi_{\rm min}^2 \simeq 7.76 \end{array}$ 
&& \hspace{-0.2cm}$\begin{array}{l} 3\sigma {\text{ range}} 
\\\chi_{}^2 \le 9\end{array}$ \\ \hline
		$S^{1}_{\triangle_{\mu\tau}^{}}/10^{-2}$ && $1.10$ && 
		$(0.001, 1.15)$ && $-0.002$ && $(-0.17, -0.001)$ \\
		$S^{2}_{\triangle_{\mu\tau}^{}}/10^{-2}$ && $0.69$ && 
		$(0.001, 0.74)$ && $-0.002$ && $(-0.18, -0.001)$ \\
		$S^{3}_{\triangle_{\mu\tau}^{}}/10^{-3}$ && $-0.85$ && 
		$(-0.92, -0.001)$ && $0.002$ && $(0.001, 0.18)$ \\
		\hline 
		$\phi_{\mu 1}^{\prime}-\phi_{\tau 1}^{\prime}$ && $4.23^{\circ}$
		&&$(0.001^{\circ},4.31^{\circ})$ && $-0.002^{\circ}$ && 
		$(-0.15^{\circ},-0.0004^{\circ})$  \\
		$\phi_{\mu 2}^{\prime}-\phi_{\tau 2}^{\prime}$ && $-2.67^{\circ}$
		&&$(-2.82^{\circ},0.13^{\circ})$ && $-0.001^{\circ}$ 
		&& $(-0.18^{\circ},-0.001^{\circ})$  \\
		$\phi_{\mu 3}^{\prime}-\phi_{\tau 3}^{\prime}$ &&$-1.56^{\circ}$ 
		&& $(-1.65^{\circ},-0.002^{\circ})$ && $0.003^{\circ}$ 
		&& $(0.003^{\circ},0.31^{\circ})$ 
		\\  \hline 
		${\cal{J^{\prime}}}/10^{-2}$&& $-3.29$  && $(-3.48,-3.09)$ && $-3.32$ && 
		$(-3.39,-3.24)$
		\\
		\hline
		$m_1^{\prime}$ or $m_3^{\prime}$/eV && $1.4\times 10^{-5}$  
		&& $(0,0.1)$ 
		&& $0.098$ && 	$(0,0.1)$ \\ \hline
		$\tan\beta$&& $50$  && $(10,50)$ && $10$ && 
		$(10,50)$ \\
		\hline\hline
	\end{tabular}
	\vspace{0.2cm}
\end{table}
\begin{figure}[h t p]
	\caption{The illustration of six LUTs of the Dirac neutrinos
	at $\Lambda_{\rm EW}^{}$ and $\Lambda_{\mu\tau}^{}$ in 
	the complex plane,
    corresponding to the best-fit values of the NMO case 
    in Table 2 $(\chi^2_{\rm min}\simeq 0.01)$,
    where the LUTs at $\Lambda_{\rm EW}^{}$ $(\Lambda_{\mu\tau}^{})$ 
    are plotted in blue (black) color,
    and the notations of sides and inner angles
    belong to the blue LUTs at $\Lambda_{\rm EW}^{}$.}
	\vspace{0.6cm}
	\centering
	\includegraphics[]{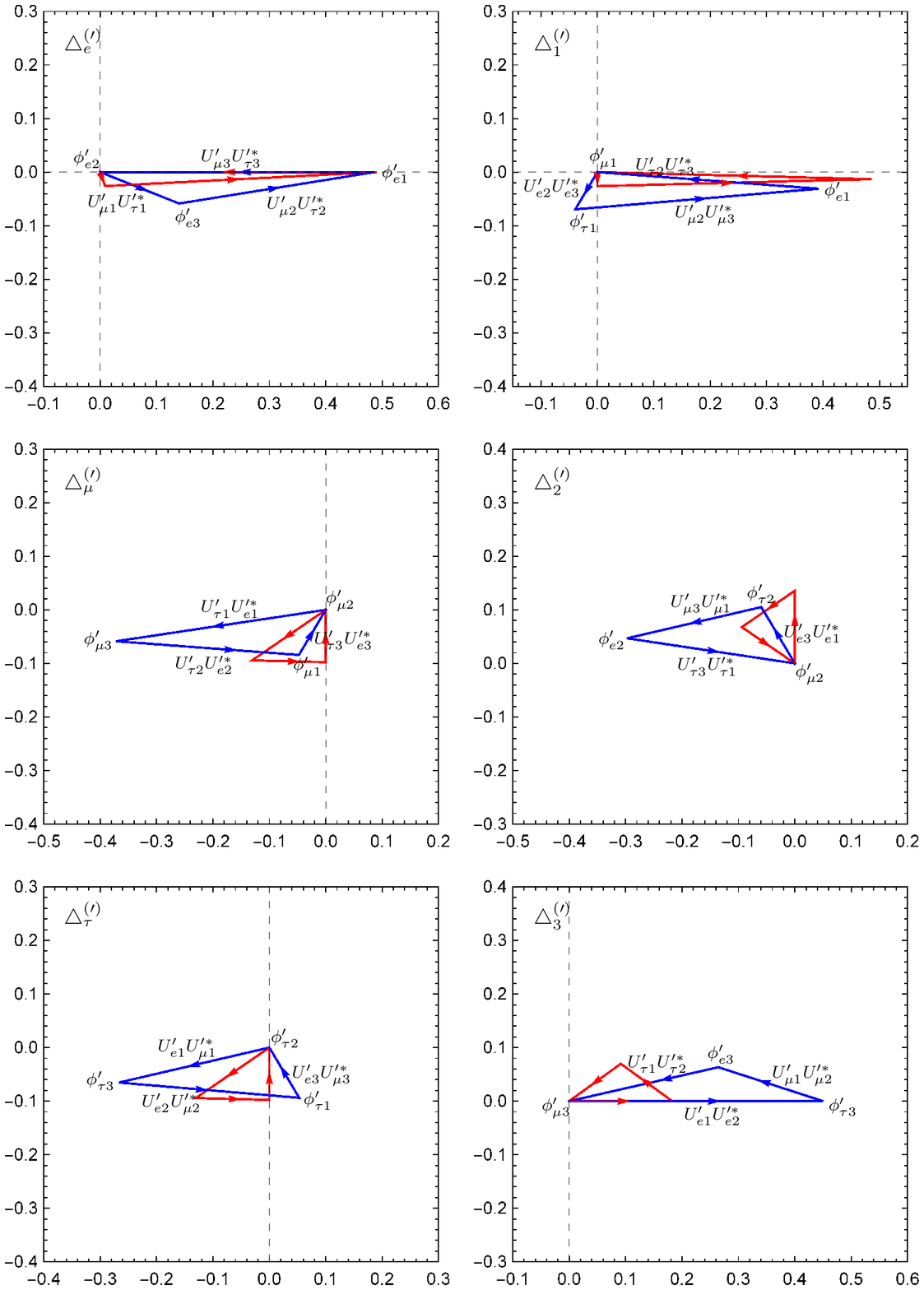}
\end{figure}
\begin{figure}[h t p]
	\caption{The illustration of six LUTs of the 
		Majorana neutrinos
		at $\Lambda_{\rm EW}^{}$ and $\Lambda_{\mu\tau}^{}$ in 
		the complex plane,
		corresponding to the best-fit values of the NMO case 
	    in Table 5 $(\chi^2_{\rm min}\simeq 0.01)$,
		where the LUTs at $\Lambda_{\rm EW}^{}$ $(\Lambda_{\mu\tau}^{})$ 
		are plotted in blue (black) color,
		and the notations of sides and inner angles
		belong to the blue LUTs at $\Lambda_{\rm EW}^{}$.}
	\vspace{0.6cm}
\centering
\includegraphics[]{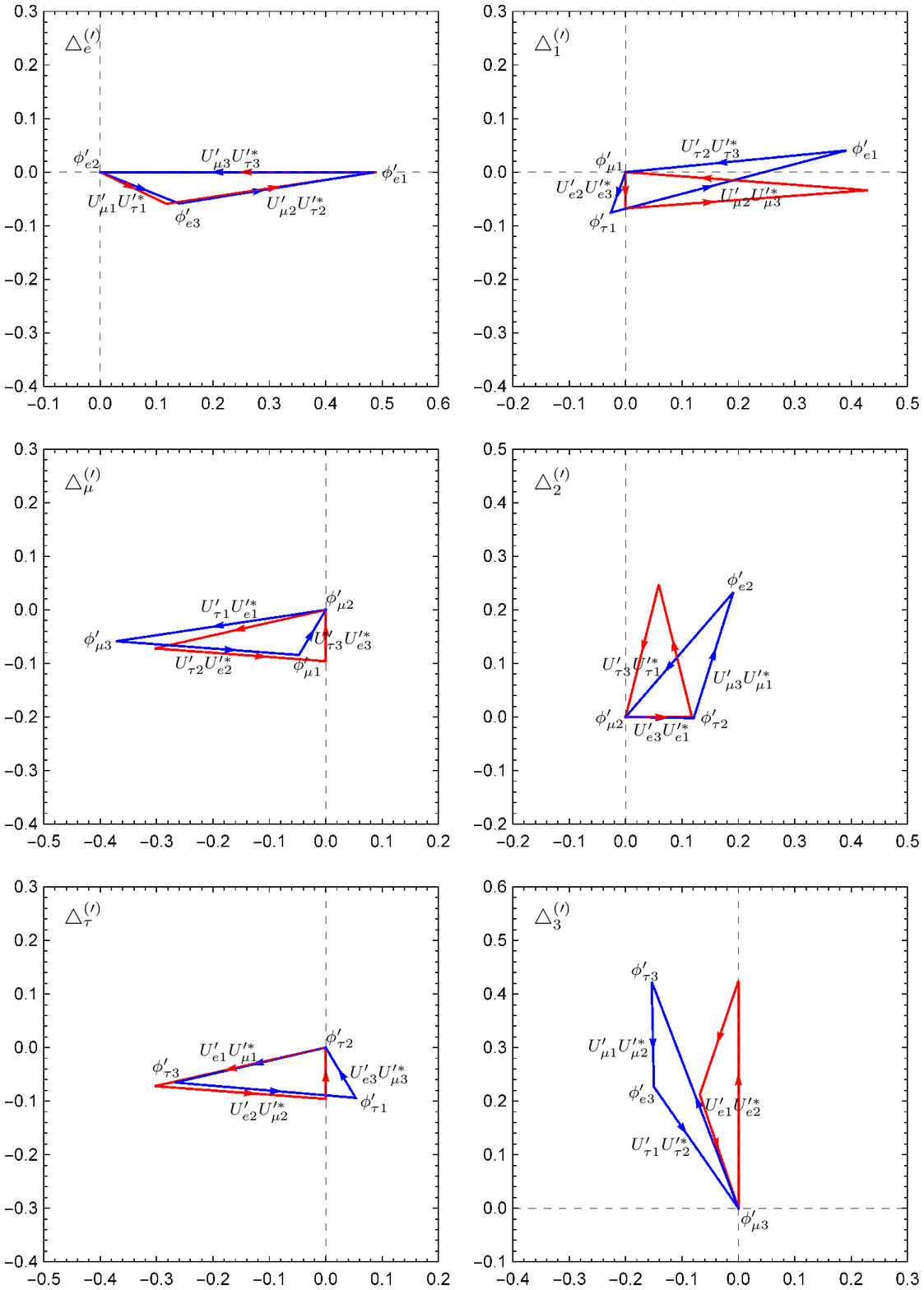}
\end{figure}
\section{Summary}
The neutrino physics has promisingly enteblack the era of precision measurements,
providing us more information to understand the large-angle
lepton favor mixing pattern and potentially big CP-violating phases.
From the perspective of model construction, we usually introduce
heavy degrees of freedom and flavor symmetry at a superhigh energy scale
to explain the smallness of neutrino masses and the observed results of
neutrino oscillation parameters at $\Lambda_{\rm EW}$. 
In this paper, we use the LUTs to 
describe the RGE running effects of lepton flavor mixing. The
analytical results in the integral form can directly connect
two LUTs at $\Lambda_{\rm H}^{}$ and $\Lambda_{\rm EW}^{}$, 
and they complement to the corresponding results of the 
differential form in Ref. \cite{Luo}. 
We also apply the LUT language to the description of the $\mu$-$\tau$ reflection symmetry, 
whose RGE-induced breaking effects can be intuitively interpreted
as the deviations of the LUTs from their special shapes at $\Lambda_{\mu\tau}^{}$.
The reformations of the six LUTs from $\Lambda_{\mu\tau}^{}$ to
$\Lambda_{\rm EW}^{}$ have been analytically and numerically studied in
a general way. Their dependence on the lightest neutrino mass, 
neutrino mass ordering, Majorana phases and the MSSM parameter
$\tan\beta$ have been revealed, corresponding to the dependence
of specific flavor mixing parameters on these factors
\cite{Xing:2015fdg,Huang,Xing:2017mkx,mutauRGE,Nath:2018hjx,Zhou:2014sya}. 
We hope this work can
enrich the neutrino phenomenology and help to understand
the relevant underlying physics.

\section*{Acknowledgements}	
I would like to thank Prof. Zhi-zhong Xing for suggesting me to
study this topic, many useful discussions and reading
the manuscript.
I am also grateful to Di Zhang and Guo-yuan Huang for 
many useful discussions. This
work was supported in part by the National Natural
Science Foundation of China under grant No. 11775231.
\appendix
\section{The exact expressions of $m_i^{\prime 2}$}
By solving Eq. (11), 
we get the exact expressions of Dirac neutrino mass
squares 
$m_i^{\prime 2}$ at $\Lambda_{\rm EW}^{}$ and
write them as
\begin{eqnarray}
m_1^{\prime 2} \hspace{-0.17cm} & = &\hspace{-0.17cm}
\frac{x}{3} -\frac{\sqrt{x^2 - 3y}}{3} \left[z + \sqrt{3\left(1-z^2\right) 
}\right] \;, \nonumber\\
m_2^{\prime 2} \hspace{-0.17cm} & = &\hspace{-0.17cm}
\frac{x}{3} -\frac{\sqrt{x^2 - 3y}}{3} \left[z - \sqrt{3\left(1-z^2\right) 
}\right] \;, \nonumber \\
m_3^{\prime 2} \hspace{-0.17cm} & = &\hspace{-0.17cm}
\frac{x}{3} +\frac{2 z \sqrt{x^2 - 3y}}{3} \;,
\end{eqnarray}
where $x=b$, $y =\left(b^2 -c\right)/2$  and
\begin{eqnarray}
z \hspace{-0.17cm} & = &\hspace{-0.17cm}
\cos \left[\frac {1}{3} \arccos \frac{2
x^3 - 9 x y + 27 a}{2 \left(x^2 - 3 y\right)^{3/2}}\right].
\end{eqnarray}
The Majorana neutrino mass squares
$m_i^{\prime 2}$ at $\Lambda_{\rm EW}^{}$
can also be exactly shown as the same form of Eq. (53) by 
replacing the definitions of $(a,b,c)$ with those defined 
below Eq. (29).
Note that Eq. (53) applies to the NMO case and in the
IMO case, we need to do the replacements $m_1^{\prime 2} \to 
m_3^{\prime 2}$, $m_2^{\prime 2} \to m_1^{\prime 2}$
and $m_3^{\prime 2} \to m_2^{\prime 2}$.
 One can find that the three Dirac or 
Majorana neutrino mass squares $m_i^{\prime 2}$
at $\Lambda_{\rm EW}^{}$ running from $\Lambda_{\rm H}^{}$ 
look similar to the effective neutrino mass squares in
constant density matter \cite{Barger:1980tf}
except the different expressions of $x$, $y$ and $z$ therein. 

\end{document}